\documentclass[twocolumn,preprintnumbers,amsmath,amssymb]{revtex4}

\usepackage{graphicx}% Include figure files
\usepackage{dcolumn}% Align table columns on decimal point
\usepackage{bm}% bold math
\usepackage{cancel}

\usepackage{amsmath,mathtools}%ams math
\usepackage{amsfonts}%ams fonts
\usepackage{amssymb}%ams symbols

\newcommand{\beq}{\begin{equation}}
\newcommand{\ber}{\begin{eqnarray}}
\newcommand{\eeq}{\end{equation}}
\newcommand{\eer}{\end{eqnarray}}

\begin{document}

\preprint{SUNY BING 9/9/2019}

\title{Generalized Ray Spaces for Paraparticles}
\author{Charles A. Nelson}
\email{cnelson@binghamton.edu}

\affiliation{%
Department of Physics, State University of New York at Binghamton, Binghamton, New York 13902-6016\\
}%

\date{ September 9, 2019 } % It is always \today, today,

\begin{abstract}
 Paraparticles of order $p = 2$ must be pair produced, so the least massive are absolutely stable.  Consequently, paraparticles are excellent candidates to be associated with dark matter and/or dark energy.   For a fixed number of paraparticles, in a ``generalized ray space" there are simple orthonormal bracket ``ray representatives" constructed from paraparticle creation operators.  Each such ray representative is associated with a Young Diagram with a unique $  {\overline P} _{sum} $ eigenvalue.   $  {\overline P} _{sum} $  is the sum of the particle-exchange operators $ \overline{P}_{i,j} $ which exchange the $i$ and $j$ identical particles.  In this paper, by using totally symmetric and totally antisymmetric brackets, generalized ray spaces for order $p=2$ are constructed for parabosons (pB), for parafermions (pB), and for possible para-Families (pFam).  Para-Families occur in the case of non-commuting pB and pF creation operators.    Explicit arbitrary order $p$ ray representatives for a few paraparticles may be useful to establish the order $p$ if, indeed, dark matter and/or dark energy is discovered to exhibit parastatistic properties.  Up to $p$  identical pB's (pF's) can occupy a totally antisymmetric (symmetric) state, unlike for normal boson (fermion) statistics.

\end{abstract}

 %\pacs{Valid PACS appear here}% PACS, the Physics and Astronomy1	
\maketitle

\section{Introduction}

It will be information on the second-quantized properties of particles associated with dark matter and/or dark energy which will be required to exclude or support the hypothesis that these new phenomena are associated with identical particles obeying parastatistics [1-5].  Information will come from data from particle astrophysics, from high energy collider experiments, and from neutrino physics. The fermions and bosons of the standard model of particle physics are order $p=1$ parastatistical quanta.   The second-quantized ray representations of bosons (fermions) occur associated with Young Diagrams with a  totally symmetric row (a totally antisymmetric column) which are 1-dimensional representations of the permutation group.  Only relative phases of these quantum states are measurable; the overall phase is not.

For $p \geq 2$ parastatistics, higher dimensional ray representations of the permutation group occur called ``generalized rays." In parastatistics, these ``generalized rays" are the fundamental physical units in a quantum domain of single mode occupancy in which individual paraparticles cannot be distinguished.    For an arbitrary number of paraparticles of order $p=2$, the generalized rays for single mode occupancy, are treated in this paper.  Their bracket ray representative states, which can be defined using the second quantization formalism, are the physical external initial and final states to be used in perturbative S-matrix-element calculations so that the identical paraparticles in the asymptotic states manifestly obey parastatistics.  The parastatistic's quanta in these external states are represented in the permutation group basis [5].  Indeed, the identical quanta in a such ray are not classically independent.   Instead, they are explicitly defined by non-commuting, non-anticommuting second-quantized creation operators through trilinear commutation relations. These quanta for $p \geq 2$ are more intricately correlated in higher dimensional rays than in the case of the 1-dimensional rays which occur for ordinary bosons and fermions.  

When useful, the paraparticles/parafields are denoted by a ``breve" accent to distinguish them from $p=1$ ones [5].

By second-quantization, for parabosons (parafermions) each generalized ray is uniquely associated with a distinct Young diagram for the permutation group.    For arbitrary order $p$ parabosons (pB), the generalized ray's Young diagram has at most $p$ rows, but never more boxes in a lower row than in a higher row.  
For arbitrary order $p$ parafermions (pF), the generalized ray's Young diagram has at most $p$ columns, but reading from the left, never more boxes in a column farther to the right.  
These irreducible ``generalized rays" are invariant under the particle-exchange operators $ \overline{P}_{i,j} = \overline{P}_{j,i}$ which exchange the $i$ and $j$ identical paraparticles [6].   For $n\geq 3$ identical paraparticles, these $ \overline{P}_{i,j}$ are non-commuting.

These  ``permutation group basis" generalized rays will occur in a scattering or decay process with a final state (initial state) having a fixed number $n$ of identical pB's and/or $\nu$  identical pF's [5].  The ray will have a unique $  {\overline P} _{sum} $ eigenvalue for an asymptotic state with all pB's, and a unique $ {\overline {\mathcal P}} _{sum}$ eigenvalue for one with all pF's.   When the pB and pF operators commute, there are a unique pair of $  {\overline P} _{sum} $ and $ {\overline {\mathcal P}} _{sum}$ eigenvalues for the pB's and pF's rays in the asymptotic state.  In the case of a para-Family (pFam) the generalized ray for the asymptotic state is partially specified (see below) by a  $  {\overline P} _{sum} $ eigenvalue for an $n$-number of pB's and a $ {\overline {\mathcal P}} _{sum}$ eigenvalue for a $\nu$-number of pF's.   

Since the particles are identical and their permutations are specified by the permutation group, parastatistics requires the ``permutation group basis" (PGB) to be used for the external states for physical measurements for pB's and similarly for pF's.  However, the ``occupation number basis" (ONB) will also be used in this paper.  The occupation number basis was used in [5], in the intermediate stages, in the calculation of the contribution of the usual covariant perturbative diagrams to the decay amplitude $ - i \mathcal{M} $ with the permutation basis used for specifying the paraparticles in the physical asymptotic states.  The superposition principle for second quantized states allows this pragmatic usage of both these bases in parastatistics.  

In this paper, using totally symmetric and totally antisymmetric brackets, three generalized ray spaces for order $p=2$ are successively  constructed:  for pB's, for pF's, and for possible pFam's of non-commuting pB and pF operators.  When pB and pF operators commute (or anticommute), the generalized ray space is a direct product of the pB and pF ones.  For each generalized ray, there is an explicit ``generalized ray representative" in a bracket form.  For $p=2$ the successive operator terms in the sum of terms for the explicit ray representative are not orthonormal.  However, each explicit generalized ray representative itself is orthonormal in the entire generalized ray space and versus the other ray representatives in its ray multiplet.

Perhaps pB and pF creation and annihilation operators for $p=2$ simply commute (or anticommute) as occurs for ordinary bosons and fermions.  However, empirical information may show that pB's and pF's are non-commuting because these new quanta occur in para-Families.   For a pFam, since the pB's and pF's are not identical particles, one does not define and use particle exchange operators to exchange a pB and a pF.  However, the trilinear commutation relations which occur for a pFam for $p \geq 2$ do have distinctive consequences for the asymptotic states.  The permutation group ray representatives in the case of pFam's are more complex than a simple direct product, but correspondingly have richer empirical signatures.    The particle exchange operators  $ \overline{P}_{i,j}$ for pB's and  $ \overline{\mathcal P}_{\iota,\kappa}$ for pF's commute irrespective of whether or not the quanta occur in pFam's.

Mathematically, this occurrence of para-Families in second-quantized quantum field theory is indicative of the fact that second-quantization enables not only a ``quantum realization" of the classical permutation of identical objects, but also generalizes the classical ``identity" concept.   In parastatistics, in many ways there is more mathematically than just a classical treatment of the permutation group.   Obviously, the second-quantized treatment in parastatistics of multi-particle occupancy of  modes goes conceptually beyond a classical  physics permutation of identical objects.  The second quantized formulation of the ``identity of non-commuting quanta" in a single mode is possible in parastatistics, whereas a classical formulation is not. 

Throughout $p=1$ physics, from many-body and condensed matter systems, through nuclei, and through hadronic physics, pairing phenomena via a condensate or otherwise can be dynamically dominant.  Unlike the occurrence of a single ray pair for two ordinary bosons or two fermions, for $p\geq 2$ there always occur two ``$p$-pairs" for 2 pB's, 2 pF's, or a pFam of 1 pB and 1 pF.    One ray-pair is symmetric and the other antisymmetric under exchange of the constituent quanta.  There is a single ray pair when there is not a pFam but instead commuting pB and pF creation and annihilation operators.
  
Anticipating that both paraparticle ``$p$-pairs" will be important, which might be occurring associated with dark energy, we treat properties associated with these two ``$p$-pairs" throughout this paper.   By the single mode labeling of the quanta, the two quanta in the ``$p$-pair" can both be particles (antiparticles) or one can be a particle and the other an antiparticle, or one can have one helicity and the other the same or the opposite helicity, etc..

Section II contains the explicit parabose bracket ray representatives for a small number of pB's and for an arbitrary $n$ number of pB's.  There are associated equations such as for the  $  {\overline P} _{sum} $ eigenvalues $e(r)$ for the $r$th ray in the ray multiplet.  Section III contains the analogous ray representatives for all parafermions.  For second quantized systems with both pB's and pF's, Section IV treats both the simple direct product case when pB and pF creation operators commute, and treats the more complex but richer case of a para-Family.   For a pFam, each of the four Kinds ray multiplets naturally occurs in its analogous more inclusive ``Meta-Multiplet."  For arbitrary numbers of pB's and pF's, the bracket ray representative in the case of a pFam has a sandwich-like operator form.  

While most of the results in this paper are for $p=2$ order parastatistics, some of the results generalize to arbitrary $p$.  In Section V, explicit orthonormal ray representatives for a few paraparticles are listed which may be useful in establishing the order $p=2$ if paraparticles occur in nature.  Section VI has ``exchange relations" for the brackets introduced in constructing ray representatives.  Section VII has some summary remarks.

In Appendix A, for the reader's reference, there is a list of the fundamental trilinear commutation relations for arbitrary order $p$ in a compact form, and a list in the conventional form.  The $p=2$ trilinear commutation relations in conventional form are listed in Appendix A of [5].  Appendix B reviews aspects of the fundamental Green indices, in particular their importance for constructing arbitrary order $p$ orthonormal generalized rays and for deriving the ``exchange relations for the brackets" in Section VI.
 
\section{Paraboson Generalized Ray Space \newline for order $p=2$}

Through out this paper, in listing the ``bracket ray representative" states, we are suppressing an extra $\frac{1}{\sqrt{2}}$ state-normalization factor for each paraparticle creation operator.     These factors arise for consistency with ``arbitrary $p$ normalization" of [2-4], see first footnote of [5]. When these extra factors are included, each listed ray representative is normalized.

For parabosons, the other rays in a higher $ d$-dimension Young diagram for $p \geq 2$ are
 generated from the ``ray representative" associated with the Young diagram $ Y^{ l } _n$  by acting with the particle-exchange operators $ \overline{P}_{i,j} $ for identical parabosons.  These  $ \overline{P}_{i,j} $ are non-commuting for $n \geq 3$. 
 
 The subscript $n$ is the number of parabosons in the generalized ray.  $n$ is the number of boxes in the associated Young diagram with $l=$(number of boxes missing in second row of the Young diagram, versus those in first row).  $n$ is the number of creation operators in each term in the bracket ray representative.
 
 For $p=2$ the Young diagram has only two rows of boxes, and the number in the second row is at most as many as are in the first row.  For the $n$-particle paraboson ray multiplet $   \overline Y^{} _n $  there are 
 \ber 
  n_{\overline{ P} }=\frac{1}{2} n(n-1)     \nonumber
 \eer
 particle-exchange operators which is also the $  {\overline P} _{sum} $ eigenvalue of its 1-dimensional ray $ Y^{n} _n $.  
 For 3 parabosons, 
\ber
 {\overline P} _{sum} = \overline{P}_{a,b}+\overline{P}_{b,c}+\overline{P}_{c,a}.   \nonumber
\eer
The various ray subspaces $ Y^{ l } _n$ of the ray multiplet 
 \newline
 $   \overline Y^{} _n  $ are each  
  invariant under the action of these particle-exchange operators $ \overline{P}_{i,j}$.
  
For systems with $n$ parabosons, for $p=2$  the associated ray multiplet  is
\ber
\overline{Y}^{ l_{min} } _n = \{ Y^{n}_n,Y^{n-2}_n, \cdots,Y^{l_{min}-2}_n,Y^{l_{min}}_n        \}
\eer
with the successive rays labeled by $r=1, \cdots , r_{max}$; $r=\frac{1}{2} (n-l+2)$.
It contains 
\ber
r_{max}= \frac{1}{2} ( n- {l_{min}} +2 ) 
\eer
 rays, where ${l_{min}}$ is for the last $r_{max}$ ray in the multiplet.  For that last ray ${l_{min}}=$(number of boxes missing in second row of the Young diagram, versus those in first row).  
 For $n=$ even-valued, ${l_{min}}=0$ and for $n=$ odd-valued, ${l_{min}}=1$.

The $r$-th ray, ${Y}^{ l } _n$,  is uniquely specified by the number of parabosons $n$ and  by its $  {\overline P} _{sum} $ eigenvalue $e(r)$.   Usually the ray is uniquely specified by its Young diagram's dimension $d(r) =$(number of states in  ${Y}^{ l }_n$).  For $n=6, 7$, there are 2 rays with the same number of states.

\subsection{\label{sec:level2} Paraboson Bracket Ray Representatives}

The ray multiplet $ \overline Y^{ } _n =  \overline Y^{} _2 $  for $n=2$ parabosons has two 1-dimensional irreducible representations  $ Y^{ l } _n = Y^{ 2 } _2 $ and $ Y^{ 0} _2 $ with respective bracket ray representatives
\ber
\frac{1}{\sqrt{2}}      \{  {a}^{\dag} ,{b}^{\dag}  \}       |0> ,
\frac{1}{\sqrt{2}} [{a}^{\dag} ,{b}^{\dag} ] |0>
\eer    
These ray representatives correspond to the 2-box totally symmetric row (antisymmetric column) Young diagrams.   This  ``minimal labeling" $ Y^{ l } _n $ gives simply the explicit bracket form for the ray representative and its associated Young diagram.  

For arbitrary order $p$ parabosons, the $ Y^{ \{ l_i \}} _n$ ray representatives' Young diagram has at most $p$ rows, but never more boxes in a lower row than in a higher row.  The superscript $l$ is replaced by a set $\{ l_i \}$ where $l_i=$(number of boxes missing in $i$th row of the Young diagram, versus those in the row above it) with $i=1,\cdots (p-1)$.   

A fuller labeling for $p=2$ is sometimes useful:  The ``full labeling" of a parabose ray representative is $ \prescript{}{d} Y^{ l } _{n,e}$. 
The $  {\overline P} _{sum} $ eigenvalue $e$ these two rays in (3) is respectively $ \pm 1$.  To label the parabose ray representatives with this eigenvalue a following subscript $e$ is added and to include the dimension of the associated ray space a pre-subscript $d$ is added.     With this full labeling, the above two ray representatives are  $ Y^{ 2 } _2=\prescript{}{1} Y^{ 2} _{2,1}$, $Y^{ 0} _2=\prescript{}{1} Y^{ 0 } _{2,-1}$.

The specific Roman (Greek) letter for parabosons (parafermions) corresponds to a specific mode index $i,j,k \cdots$ which can be denoted by a subscript, ${a}^{\dag}_{i}$.  This index includes the momentum components, and the helicity components for the para-Majorana neutrino field $ \breve{\xi}=\breve{\nu}$, and  the $\breve{A}$, $\breve{B} \equiv  \bar{\breve A}$ particle-antiparticle distinction for the $\breve{\mathcal{A}}$ complex field considered in the supersymmetric ``statistics portal" Lagrangians considered in [5].   These same equations apply for para-Dirac fields, for real scalar parafields, etc. of order $p=2$.

In the following bracket ray representatives, a pB bracket $ \{a_1, \cdots , a_n \}$ with $n$-operators is totally symmetric.  A pF bracket $[a_1, \cdots , a_n]$ is totally antisymmetric.  Each resultant bracket ray representative is independent and orthogonal.   The  ``flip" reordering relations  [7] for parabosons (parafermions).
\ber
{a}^{\dag} {b}^{\dag}  {c}^{\dag}   =   {c}^{\dag}  {b}^{\dag} {a}^{\dag}    \nonumber \\
{\alpha}^{\dag} {\beta}^{\dag}  {\gamma}^{\dag}   =   - {\gamma}^{\dag}  {\beta}^{\dag} {\alpha}^{\dag}    
 \eer
are used so these pB and pF brackets are a sum of independent terms. These useful relations are the ``self-contained trilinear commutation relations" of $p= 2$ parastatistics for all creation/annihilation operators (see Appendix A).  They also hold for ordinary bosons and fermions.

Below in Section V, when we treat the case of arbitrary $p$ ray representatives, a ``prime" superscript will be added to the analogous totally symmetric (totally antisymmetric) brackets to indicate that in those brackets these ``flip" reordering relations have not been used.   For a para-Family, the plus (minus) sign in (4) occurs when respectively there are more pB (pF) operators, see (37-38).

Due to (4), these $p=2$ brackets have fewer independent terms than the ``primed" ones.  For instance,
\ber
 \{  {a}^{\dag} ,{b}^{\dag} , {c}^{\dag} \} = ({a}^{\dag} {b}^{\dag} {c}^{\dag} + {b}^{\dag} {c}^{\dag} {a}^{\dag} +{c}^{\dag} {a}^{\dag} {b}^{\dag} )  \nonumber\\
 = \frac{1}{2} \{  {a}^{\dag} ,{b}^{\dag} , {c}^{\dag} \} ^{'}
\eer
A  ``primed" bracket has $n!$ independent terms.  The number of independent terms in the unprimed pB/pF bracket is the same as the dimension $\overline {d}_n$ ($\overline {\delta}_{\nu}$)of the associated ray multiplet, $ {\overline Y^{ } }_n $  ($   \overline {\mathcal Y}^{} _{\nu} $).  For parafermions  the minus signs of a ``primed bracket" can be absorbed in the pF bracket,
\ber
[ {\alpha}^{\dag} ,{\beta}^{\dag} , {\gamma}^{\dag} ]= ({\alpha}^{\dag} {\beta}^{\dag} {\gamma}^{\dag} + {\beta}^{\dag} {\gamma}^{\dag} {\alpha}^{\dag} +{\gamma}^{\dag} {\alpha}^{\dag} {\beta}^{\dag} )  \nonumber\\
 = \frac{1}{2} [ {\alpha}^{\dag} ,{\beta}^{\dag} , {\gamma}^{\dag} ] ^{'}
\eer

These ``flip" reordering relations (4) for $p=2$ preserve the even/odd place position [6] of the successive operators whether reading from left or right, unlike for $p \neq 2$ parastatistics.  This even/odd place-position symmetry holds in the fundamental $p=2$ trilinear relations for pB, pF, and pFam paraparticles (see Appendix A in [5]).   It is a symmetry useful in parts of a operator analysis.  However, for $p=2$, this place-position symmetry is broken by these bracket generalized ray representative states.  It is also broken by the bracket-quantization-procedure for para-Lagrangians [2, 8, 4, 5].  Hence this even/odd place-position symmetry is broken in the contribution of a perturbative diagram to $ - i \mathcal{M} $ where $\mathcal{M}$ is the covariant amplitude.

For $3$ parabosons, the ray multiplet $   \overline Y^{} _3 $ has two rays  $ Y^{ 3 } _3 = \prescript{}{1} Y^{ 3} _{3,3}$, $  Y^{ 1 } _3 = \prescript{}{2} Y^{ 1 } _{3,0}$ with respective ray representatives
\ber
\frac{1}{\sqrt{3}}      \{  {a}^{\dag} ,{b}^{\dag} , {c}^{\dag} \}       |0> ,
\frac{1}{\sqrt{2}} [{a}^{\dag} ,{b}^{\dag}] {c}^{\dag}|0>
\eer

For $4$ parabosons, the ray multiplet $   \overline Y^{} _4 $   has three rays  $ Y^{ 4 } _4 = \prescript{}{1} Y^{ 4} _{4,6}$, $  Y^{ 2 } _4= \prescript{}{3} Y^{ 2 } _{4,2}$, $  Y^{ 0} _4= \prescript{}{2} Y^{ 0 } _{4,0}$ with respective ray representatives
\ber
\frac{1}{\sqrt{6}}      \{  {a}^{\dag} ,{b}^{\dag} , {c}^{\dag} , d^{\dag}\}       |0> ,
\frac{1}{2}[ {a}^{\dag} ,{b}^{\dag} ]  \{ {c}^{\dag} ,{d}^{\dag} \}|0>, \nonumber \\
\frac{1}{2} [{a}^{\dag} ,{b}^{\dag}]  [{c}^{\dag} ,{d}^{\dag}]|0>
\eer
For Young diagrams symmetric about the diagonal the   $  {\overline { P}} _{sum} $ eigenvalue is zero.

As occurs in general,  the individual terms in the explicit ray representatives (7) and (8) are not orthonormal.  However, because of the explicit summation of the terms, each bracket generalized ray representative is orthonormal versus the other rays in its ray multiplet and in the entire generalized ray space.  
[Caution:  For normalization, the suppressed state-normalization factors of $\frac{1}{\sqrt{2}}$ for each creation operator must be included.]

For $5$ parabosons, the ray multiplet $   \overline Y^{} _5 $   has three rays  $ Y^{ 5 } _5 = \prescript{}{1} Y^{ 5} _{5,10}$, $  Y^{ 3 } _5= \prescript{}{4} Y^{ 3 } _{5,5}$, $  Y^{ 1} _5= \prescript{}{5} Y^{ 1 } _{5,2}$ with respective ray representatives
\ber
\frac{1}{\sqrt{10}}      \{  {a}^{\dag} ,{b}^{\dag} , {c}^{\dag} , d^{\dag}  ,{e}^{\dag} \}       |0> ,
\frac{1}{\sqrt{6}}[ {a}^{\dag} ,{b}^{\dag} ]  \{ {c}^{\dag} ,{d}^{\dag} ,{e}^{\dag} \}|0>, \nonumber \\
\frac{1}{2} [{a}^{\dag} ,{b}^{\dag}]  [{c}^{\dag} ,{d}^{\dag}]   {e}^{\dag} |0>
\eer

These cases display the general pattern in the bracket representative for each allowed $p=2$ Young diagram.  Note that the successive columns of the Young diagram are alphabetically filled with Roman letters as one moves across, with each 2-box column corresponding to a $[ {a}^{\dag} ,{b}^{\dag} ]$ antisymmetric bracket.  This product of successive 2 particle brackets is then followed by
 an $l$-length totally-symmetric bracket, such as $ \{   {c}^{\dag} , d^{\dag}  ,{e}^{\dag},{f}^{\dag} ,{g}^{\dag} \}    $ for $l=5$.   
 
 In the alternate, general notation introduced later in this paper for the para-Family generalized ray space in Section IV, see (53):  
 
 The generalized ray representative $ Y^{ l } _n $ for an Even number $n_E$ of pB's is 
 \ber
 | r  > |_{n_E} = \prod^{h_{max}}_{h=2 } ( \frac{1}{\sqrt{2}} [ a^{\dag}_{h-1}, a^{\dag}_h] )  \nonumber \\
  \frac{1} { \sqrt{ {\overline d}_{l_E} }  }  \{ {b^{\dag} }_{h_{max} +1 },  \cdots ,{b^{\dag}_{n_E}} \}  |0>  
\eer
The mode subscript on the pB creation operators includes both the momentum components and the particle-antiparticle $\breve{A}$, $\breve{B}= \overline{\breve{A}}$ distinction for a $\breve{\mathcal{A}}$ complex spin-0 field.  For a higher spin pB field, this subscript will also include the polarization components.

The even valued $l_E = n - 2(r-1)=2(r_{max} -r) $,  is the length of the totally symmetric bracket; with $r_{max} = \frac{n} {2} +1$.  It's normalization is 
\ber
 {\overline d}_{l_E} =  \frac{l !}{(\frac{l}{2}) !(\frac{l}{2} )!} 
\eer
The number of pB operators in the commutator factors on the left is
\ber
h_{max}=n- l = 2r-2 ,
\eer

For $ | r  > |_{n_O}$ with an Odd number $n_O$ of pB's: Replace $n_E$ by $n_O$ in (10).  The odd valued $l_O = n - 2(r-1) =2(r_{max} -r) +1 $ is the length of the totally symmetric bracket; with $r_{max} = \frac{n +1 }{2}$.  In (10), replace $ {\overline d}_{l_E} $ with $ {\overline d}_{l_O} $
\ber
{\overline d}_{l_O} = \frac{l !}{(\frac{l+1}{2}) !(\frac{l-1}{2} )!} 
\eer
 
 \subsection{\label{sec:level2} Equations for Paraboson Generalized Ray Space }
 
 For each pB ray, the parabose $  {\overline P} _{sum} $ eigenvalue $e(r)$ can be used to define ``effective" numbers of symmetric and of antisymmetric pairs by
\ber
n_{s,a }= \frac{1}{2} [ ( n_{\overline{ P} } \pm e(r) ]
\eer
where $n_{\overline{ P}}$ is the number of $ \overline{P}_{i,j} $ pB particle-exchange operators.   
For the 2-particle paraboson symmetric ray $ \{n_s, n_a\} = \{1, 0 \}$ and for the antisymmetric ray $ \{n_s, n_a\} = \{0, 1 \}$.   

For totally symmetric 1-dimensional rays $n_a =0$. 
For $p=2$ both $n_{s,a }$ are non-negative, and respectively monotonically decrease (increase) as $r$ increases through the paraboson ray multiplet.

The $  {\overline P} _{sum} $ eigenvalue 
\ber
e(r) =  e(n,l) = \frac{1}{4} ( n^2 + l^2 -4n +2l)  \nonumber \\ 
= \frac{1}{2} ( {r_1}^2 + {r_2}^2 - r_1 - 3 r_2) 
\eer 
where $r_1=$(number of boxes in first row), $r_2=$(number of boxes in second row) of the $Y^{l_{}}_n$ diagram.    As $r$ increases through the ray multiplet, $e(r)$ monotonically decreases with $e(r_{max})$ vanishing only $n=3 ,4$.
\ber
n_s = \frac{1}{8} [ 3n (n-2) + l(l+2)], \nonumber \\  
n_a= \frac{1}{8} [ n (n+2) - l(l+2)]  
\eer

For a large number $n$ of parabosons, the ray with the most states, so the ``most probable ray"  $Y^{l_{mpr}}_n$, has $l_{mpr} = \sqrt{n}$, $r_{mpr}=\frac{1}{2}(n -\sqrt{n} + 2)$, is narrow in $r$, and 
\ber
e(r_{mpr}) = \frac{1}{4} [ n^2  -3n +2 \sqrt{n}],  \nonumber \\
n_s (r_{mpr})= \frac{1}{8} [ 3n^2 - 5 n  +2 \sqrt{n}], \nonumber \\  
n_a (r_{mpr}) = \frac{1}{8} [ n^2 +  n  -2 \sqrt{n}]  
\eer

For $n_{even}= n_{e}$ large, to leading order in $n$, these are the same as for the ``last ray in the multiplet," because $r_{max}=\frac{1}{2} [ n_e +2]$.  Since $n_{even}$, this last ray has a rectangular Young diagram with
\ber
e(r_{max}) = \frac{1}{4} [ {n_e}^2  -4n_e],  \nonumber \\
n_s (r_{max})= \frac{3}{8} [ {n_e}^2 - 2 n_e  ], \nonumber \\  
n_a (r_{max}) = \frac{1}{8} [ {n_e}^2 +2 n_e  ]  
\eer

The dimensions of the $Y^{l_{}}_n$ ray can be found from the ``hook value" $H(n,l)$ for its Young diagram
\ber
d(r)=d(n,l)= \frac{n!}{H(n,l)}
    =\frac{(l+1) n! }{{( \frac{n+l+2}{2})}!    ( \frac{n-l}{2})!} \nonumber \\
    =\frac{(r_1-r_2 +1) (r_1 + r_2 )! }{{(r_1 +1 )}!    ( r_2)!}  = \frac{(n- 2 r +3) n! }{{(n-r+2)}!    ( r-1)!} \nonumber \\
 \eer
 or
 \ber
 d(r)=d(n,l)=   \left( \begin{array} {c} n \\  r-1  \end{array} \right)   -    \left( \begin{array} {c} n \\  r-2  \end{array} \right) .  \nonumber 
\eer
For $n=6, 7$ there are two rays with the same dimension in the associated ray multiplet.

For the  paraboson ray multiplet $ \overline{Y}^{} _n $:   Its dimension 
\ber
\overline{d}_{ n}  = \sum\limits_{r=1}^{r_{max}}     d(r)             \nonumber
\eer
equals the number of independent states in $ \overline{Y}^{} _n $.  This ray multiplet dimension
\ber
\overline{d}_{n } = \frac{n!}{(m_o)! (m_e)!}   
\eer
where reading from the left $m_o =$( number of odd place creation operators) and $m_e =$( number of even place creation operators).  
These ray multiplet dimensions for the total number of independent states for a pB ray, or  a pF ray, can be determined in the ``occupation number basis" [5].

The dimensions of the ray multiplet also occur as the normalizations for the $n$-totally symmetric brackets
\ber
\frac{1} { \sqrt{ {\overline d}_{n} }  }  \{ a^{\dag} ,b^{\dag} ,\cdots {f_{n}}^{\dag} \} 
\eer
with $ \frac{1} { \sqrt{2} }  [ a^{\dag}, b^{\dag} ]_{\pm}   $ for 2 pB's.

For $ \overline{Y}^{} _n $, the number of rays in it is $r_{max} = m_e + 1$, and from the ``flip" SCR relations, see Appendix A, the
 [Number of independent SCR equalities]$= [(m_o)! (m_e)! -1]$.

In going from one ray multiplet to the next, there are the recursion expressions
\ber
{\overline d}_{n_e} = 2 ~ {\overline d}_{n_o = n_e -1}    \nonumber
\eer
\ber
{\overline d}_{n_o} = \frac{2 n_o}{n_o +1}  ~ {\overline d}_{n_e = n_o -1}   
\eer

The sum of the  $  {\overline P} _{sum} $ eigenvalues is
\ber 
\overline{e} (n, l_{min} ) = \sum\limits_{l=0}^{l_{min}}     e(n,l)             \nonumber \\
= \overline{e_o} (n) + \overline{e_1} (n,l_{min})
\eer
\ber
\overline{e_o} (n)=  \frac{1}{6} [ {n}^3  -4n],  \nonumber \\
\overline{e_1} (n, l_{min})= -  \frac{1}{24} [ (l_{min})^3  +  l_{min} ( 3 n^2- 3n -1) ]
\eer

\section{Parafermion Generalized Ray Space for order $p=2$}
 
For systems with $\nu$-parafermions, for arbitrary order $p$ the Young diagrams associated with generalized rays correspond very simply to a diagonal mirror reflection of rows and columns of the $n=\nu$ box parabose Young diagram.    

A parafermi generalized ray is minimally labeled $ {\mathcal Y}^{ \lambda } _\nu$.   
The subscript $\nu$ denotes the number of boxes in the parafermion Young diagram with $\lambda=$(number of boxes missing in second column of the Young diagram, versus those in first column).  For the $\nu$-particle parafermi ray there are 
 \ber 
  \nu_{\overline{ {\mathcal P}} }=\frac{1}{2} \nu (\nu-1)     \nonumber
 \eer
 pair particle-exchange operators which is also the $  { \overline{\mathcal P}} _{sum} $ eigenvalue of its 1-dimensional ray $ \mathcal Y^{\nu} _{\nu} $.    We denote the particle-exchange operators  which exchange the $\iota$ and $\kappa$ identical pF particles by $ \overline{\mathcal P}_{\iota,\kappa} =\overline{\mathcal P}_{\kappa,\iota} $.
 For 3 parafermions, 
\ber
 {\overline {\mathcal P}} _{sum} = \overline{{\mathcal P}}_{\alpha,\beta}+\overline{{\mathcal P}}_{\beta,\gamma}+\overline{{\mathcal P}}_{\gamma,\alpha}.   \nonumber
\eer
 
For systems with $\nu$ parafermions, for $p=2$ the parafermi ray multiplet is  
\ber
\overline{\mathcal Y}^{ {\lambda}_{min} } _\nu = \{ {\mathcal Y}^{\nu}_\nu , {\mathcal Y}^{\nu-2}_\nu, \cdots,{\mathcal Y}^{{\lambda}_{min}-2}_\nu,{\mathcal Y}^{{\lambda}_{min}}_\nu       \}
\eer
with the successive rays labeled by with $\rho=1, \cdots ,\rho_{max}$; $\rho=\frac{1}{2} (\nu-\lambda+2)$.  It contains
\ber
 \rho_{max}= \frac{1}{2} ( \nu- {\lambda_{min}} +2 ) 
 \eer
 rays, where ${\lambda_{min}}$ is for the last $\rho_{max}$ ray in the parafermi multiplet.      For $\nu=$ even-valued, ${\lambda_{min}}=0$ and for $\nu=$ odd-valued, ${\lambda_{min}}=1$.
 
For the parafermi ray,  a ``full labeling"  is by  $\prescript{}{\delta}{\mathcal Y}^{ \lambda } _{\nu,\epsilon} $.  It
is uniquely specified by the number of parafermions $\nu$  and by its $  {\overline {\mathcal P}} _{sum} $ eigenvalue ${
\epsilon}$.  To include the dimension of the associated ray space a pre-subscript $\delta$ is added.

For arbitrary order $p$ parafermions, the $ {\mathcal Y}^{ \{ \lambda_i \}} _n$ ray representatives' Young diagram has at most $p$ columns. But, reading from the left, never more boxes in successive column than in an earlier one.  The superscript $\lambda$ is replaced by the set $\{ \lambda_i \}$ where $\lambda_i=$(number of boxes missing in $i$th column of the Young diagram, versus those in the column to the left of it) with $i=1,\cdots (p-1)$.  

 \subsection{\label{sec:level2} Parafermion Bracket Ray Representatives}
 
 For $n=2$  parafermions, the ray multiplet $ \overline {\mathcal Y}^{ } _n =  \overline {\mathcal Y}^{} _2 $ has two rays  $ {\mathcal Y}^{ l } _n = \mathcal Y^{ 2 } _2=\prescript{}{1} {\mathcal Y}^{ 2} _{2,-1} $ and $ {\mathcal Y}^{ 0} _2=\prescript{}{1} {\mathcal Y}^{ 0} _{2,1}  $ 
\ber
\frac{1}{\sqrt{2}}      [  {\alpha}^{\dag} ,{\beta}^{\dag}  ]       |0> ,
\frac{1}{\sqrt{2}} \{{\alpha}^{\dag} ,{\beta}^{\dag} \} |0>
\eer
For $3$ parafermions, $   \overline {\mathcal Y}^{} _3 $ has two rays  $ {\mathcal Y}^{ 3 } _3 = \prescript{}{1} {\mathcal Y}^{ 3} _{3,-3}$, $  {\mathcal Y}^{ 1 } _3 = \prescript{}{2} {\mathcal Y}^{ 1 } _{3,0}$ 
\ber
\frac{1}{\sqrt{3}}      [ {\alpha}^{\dag} ,{\beta}^{\dag} , {\gamma}^{\dag} ]       |0> ,
\frac{1}{\sqrt{2}}\{{\alpha}^{\dag} ,{\beta}^{\dag}\}{\gamma}^{\dag}|0>
\eer
For $4$ parafermions,  $   \overline {\mathcal Y}^{} _4 $   has three rays  $ {\mathcal Y}^{ 4 } _4 = \prescript{}{1} {\mathcal Y}^{ 4} _{4,-6}$, $ {\mathcal Y}^{ 2 } _4= \prescript{}{3} {\mathcal Y}^{ 2 } _{4,-2}$, $  {\mathcal Y}^{ 0} _4= \prescript{}{2} {\mathcal Y}^{ 0 } _{4,0}$ 
\ber
\frac{1}{\sqrt{6}}      [  {\alpha}^{\dag} ,{\beta}^{\dag} , {\gamma}^{\dag} , {\delta}^{\dag}]       |0> ,
\frac{1}{2}\{ {\alpha}^{\dag} ,{\beta}^{\dag} \} [ {\gamma}^{\dag} ,{\delta}^{\dag} ]|0>, \nonumber \\
\frac{1}{2} \{{\alpha}^{\dag} ,{\beta}^{\dag}\} \{{\gamma}^{\dag} ,{\delta}^{\dag}\} |0>
\eer
For $5$ parafermions,  $   \overline {\mathcal Y}^{} _5 $   has three rays  $ {\mathcal Y}^{ 5 } _5 = \prescript{}{1} {\mathcal Y}^{ 5} _{5,-10}$, $ {\mathcal Y}^{ 3 } _5= \prescript{}{4} {\mathcal Y}^{ 3 } _{5,-5}$, $  {\mathcal Y}^{ 1} _5= \prescript{}{5} {\mathcal Y}^{ 1 } _{5,-2}$ 
\ber
\frac{1}{\sqrt{10}}  [  {\alpha}^{\dag} ,{\beta}^{\dag} , {\gamma}^{\dag} , {\delta}^{\dag}  ,{\epsilon}^{\dag} ]  |0> ,
\frac{1}{\sqrt{6}} \{ {\alpha}^{\dag} ,{\beta}^{\dag} \}  [{\gamma}^{\dag} ,{\delta}^{\dag} ,{\epsilon}^{\dag} ]|0>, \nonumber \\
\frac{1}{2}  \{ {\alpha}^{\dag} ,{\beta}^{\dag} \}  \{ {\gamma}^{\dag} ,{\delta}^{\dag}\}   {\epsilon}^{\dag} |0>  ~~
\eer
The general pattern is that the successive rows are alphabetically filled with Greek letters as one moves down the associated Young diagram with each 2-box row corresponding to a $ \{ {\alpha}^{\dag} ,{\beta}^{\dag} \}$ symmetric bracket .  This product of successive 2 particle brackets is then followed by
 an $\lambda$-length totally-antisymmetric bracket.   
 
In the alternate, general notation introduced later in this paper  in Section IV for the para-Family generalized ray space, see (53):  

The generalized ray representative $ {\mathcal Y}^{ l } _n $ for an Even number $\nu_E$ of pF's is 
 \ber
 | \rho  > |_{\nu_E} =  \frac{1} { \sqrt{ {\overline \delta}_{\lambda_E} }  }  [ {\beta^{\dag}_{{\nu }_E}},  \cdots ,{\beta^{\dag}_{\iota_{max} +1}}  ]  \nonumber \\
 \prod^{{\iota}_{max}}_{{\iota}=2 } ( \frac{1}{\sqrt{2}} \{ {\alpha}^{\dag}_{{\iota}}, {\alpha}^{\dag}_{\iota-1} \}) |0> 
\eer
The mode subscript on the pF creation operators includes both the momentum components and the spinor or helicity components for a para-Majorana field $\breve{\xi}$ or for a para-Dirac field $\breve{\psi}$.   

Versus the earlier examples, in the pF ray representative formula (31) the totally antisymmetric bracket is written on the left as in the pFam equation (53) below, and the pF operators are labeled from right.  Rearranging the order of the various factors can produce a change in the overall sign, see Section VI below.

The even valued $\lambda_E = \nu - 2(\rho-1)=2(\rho_{max} -\rho) $,  is the length of the totally antisymmetric bracket; with $\rho_{max} = \frac{\nu} {2} +1$.  It's normalization is 
\ber
 {\overline \delta}_{\lambda_E} =  \frac{\lambda !}{(\frac{\lambda}{2}) !(\frac{\lambda}{2} )!} 
\eer
The number of pF operators in the anticommutator factors on the right is
\ber
\iota_{max}=
\nu- \lambda = 2\rho-2 ,
\eer

For $ | \rho  > |_{\nu_O}$ with an Odd number $\nu_O$ of pF's: Replace $\nu_E$ by $\nu_O$ in (31).   The odd valued $\lambda_O = \nu - 2(\rho-1) =2(\rho_{max} -\rho) +1 $ is the length of the totally antisymmetric bracket; with $\rho_{max} = \frac{\nu +1 }{2}$.  In (31), replace $ {\overline \delta}_{l_E} $ with $ {\overline \delta}_{l_O} $
\ber
{\overline \delta}_{l_O} = \frac{\lambda !}{(\frac{\lambda+1}{2}) !(\frac{\lambda-1}{2} )!} 
\eer

 \subsection{\label{sec:level2} Equations for Parafermion Generalized Ray Space }
 
 For each pF ray, the parafermi $ {\overline {\mathcal P}} _{sum}$ eigenvalue  $\epsilon(\rho)$ can be used to define ``effective" numbers of antisymmetric and of symmetric pairs by
\ber
\nu_{a,s }= \frac{1}{2} [ ( \nu_{\overline{ \mathcal P} } \mp \epsilon(\rho) ]
\eer
where $\nu_{\overline{ \mathcal P}}$ is the number of $ \overline{\mathcal P}_{\iota,\kappa} $ pF particle-exchange operators. 
For the 2-particle parafermion antisymmetric ray $ \{\nu_a, \nu_s\} = \{1, 0 \}$ and for the symmetric ray $ \{\nu_a, \nu_s\} = \{0, 1 \}$.
For the totally anti-symmetric 1-dimensional rays $\nu_s =0$.  For $p=2$ both $\nu_{a,s }$ are non-negative, and respectively decrease (increase) as $\rho$ increases through the pararfermion ray multiplet.

Since the parafermion ray's Young Diagram $ {\mathcal Y}^{{\lambda}_{min}}_\nu $ is a diagonal flip of the corresponding paraboson ray's Young Diagram $Y^{l_{min}}_n$
 ( $\nu = n$, $\lambda_{min} = l_{min}$), the analogous equations to those in Section II for all pB's apply for all pF's by making the obvious Roman to Greek letter replacements. 
 
For instance, for a large number $\nu$ of pF's, the ray with the most states, so the ``most probable ray"   $ {\mathcal Y}^{{\lambda}_{mpr}}_\nu $, has $\lambda_{mpr} = \sqrt{\nu}$, $\rho_{mpr}=\frac{1}{2}(\nu -\sqrt{\nu} + 2)$.

 However, the $ {\overline {\mathcal P}} _{sum}$ eigenvalue  is the negative of the $ {\overline { P}} _{sum}$ eigenvalue,  $ \epsilon(\rho)  = -  e(r) $ for $\rho = r$.  Thus, $ \epsilon(\rho) \leq 0$ and $ \epsilon(\rho)  = - | e(r) |$, except for two parafermions where $\epsilon= \mp1$ respectively for totally antisymmetric (symmetric) parafermion ray-pair.

For the parafermion ray multiplet $ \overline {\mathcal Y}^{ } _n $, the dimension
\ber
\overline{\delta}_{\nu} = \sum\limits_{\rho=1}^{\rho_{max}}     \delta(\rho)             \nonumber
\eer
occurs as the normalization for the $\nu$-totally antisymmetric brackets
\ber
\frac{1} { \sqrt{ {\overline \delta}_{\nu} }   }  [ {\alpha}^{\dag} ,{\beta}^{\dag} ,\cdots {{\eta}_{\nu}}^{\dag} ]
\eer
with $ \frac{1} { \sqrt{2} }  [ {\alpha}^{\dag}, {\beta}^{\dag} ]_{\mp}   $ for 2 pF's.

\section{Para-Family Generalized Ray Space for order $p=2$}

For a second quantized system with $n$-parabosons and $\nu$-parafermions, the total number of paraparticles is denoted by $n_T = n + \nu$.

The creation and annihilation operators for pB's and pF's for $p=2$ can simply commute (or anticommute) as they do for ordinary bosons and fermions.
We label this direct product case as $ \overline{Y}_n \times  \overline{\mathcal Y}_{\nu} $, whereas we label the somewhat complex case of the pB and pF operators occuring in a para-Family by 
$ \overline{Y}_n  ~  \overline{\mathcal Y}_{\nu} $.    This overbar notation is used because the generalized rays for systems with both pB's and pF's also occur in generalized ray multiplets. 

When the pB and pF operators commute, there are a unique pair of $  {\overline P} _{sum} $ and $ {\overline {\mathcal P}} _{sum}$ eigenvalues for the pB's and pF's rays.  For the case of a para-Family (pFam) the generalized ray is partially specified by the  $  {\overline P} _{sum} $ eigenvalue for the pB's and the $ {\overline {\mathcal P}} _{sum}$ eigenvalue for the pF's.  In both cases, there are $ \frac{1}{2}[n(n-1)]$  parabose $ \overline{P}_{i,j}$ particle exchange operators and $ \frac{1}{2}[\nu(\nu-1)]$ parafermi  $ \overline{\mathcal P}_{\iota,\kappa}$ ones. The  $ \overline{P}_{i,j}$ and  $ \overline{\mathcal P}_{\iota,\kappa}$ mutually commute.

For a pFam, there are four Kinds of generalized ray multiplets  $ \overline{Y}_n  ~  \overline{\mathcal Y}_{\nu} $, each of which naturally occurs in its Kind of a more inclusive ``Meta-Multiplet" $\overline{\overline{Y} }_n   \overline{\overline{\mathcal Y}}_{\nu} $.

In general, the structures and notation occurring in this section are simple generalizations of that used above for the case of all parabosons (Section II).

\subsection{\label{sec:level2}Case: Parabosons and Parafermions Commute }

In this case the generalized ray bracket representative is the direct product $ {Y}^{ l(r) } _n \times {\mathcal Y}^{ \lambda (\rho)} _\nu $ of the respective brackets listed above.  This direct product ray occurs in the ray multiplet 
$ \overline{Y}^{ l_{min} } _n \times  \overline{\mathcal Y}^{ \lambda_{min} } _\nu $.  The ray multiplet is non-truncated in both pB's and pF's with $l_{min}$ 
equal 0 (1) when $n$ is Even-valued (Odd-valued), and analogously for $\lambda_{min}$.    The variable $r$ for labeling the sequence of rays in the ray multiplet, $r$'s range, and the formula for $l(r)$ for the ``leg" of the associated pB diagram $ {Y}^{ l(r) } _n$  are the same as for only pB's.  Likewise, for the associated pF diagram ${\mathcal Y}^{ \lambda (\rho)} _\nu $,  the pF quantities are the same as for only pF's.

\subsection{\label{sec:level2}Case: Parabosons and Parafemions Occur in Four Kinds of Para-Families}

The bracket ray representatives $| r, \rho>$ for a para-Family for $p=2$ have a sandwich-like operator form with a ``Full-Center" in the middle, pB commutators on the left, and pF anticommutators on the right.  See equation (53) below.   This operator sandwich-like format is ``generic" in that it does not depend on the Kind of pFam, but the ``Full-Center" factor does depend on which of the four Kinds.

The ``self-contained commutation relations" for a $p=2$ para-Family are trilinear relations of the creation and annihilation operators of pB's and pF's, see Appendix A of [5].  The ones which have only creation operators are ``flip" reordering relations:

For two parabosons and one parafermion:
\ber
a^{\dag}_k a^{\dag}_l {\beta}^{\dag}_m = {\beta}^{\dag}_m a^{\dag}_l a^{\dag}_k,   \nonumber \\
a^{\dag}_k {\beta}^{\dag}_l a^{\dag}_m = a^{\dag}_m {\beta}^{\dag}_l a^{\dag}_k  
\eer
For two parafermions and one paraboson:
\ber{\alpha}^{\dag}_k {\alpha}^{\dag}_l b^{\dag}_m = -  b^{\dag}_m {\alpha}^{\dag}_l {\alpha}^{\dag}_k ,   \nonumber \\
{\alpha}^{\dag}_k b^{\dag}_l {\alpha}^{\dag}_m =  -  {\alpha}^{\dag}_m b^{\dag}_l {\alpha}^{\dag}_k   
\eer
The mode index $k,l,m$ includes both the momentum components and the spinor or helicity components for a para-Majorana field $\breve{\xi}$ or for a para-Dirac field $\breve{\psi}$, and both the momentum components and the particle-antiparticle $\breve{A}$, $\breve{B}= \overline{\breve{A}}$ distinction for a $\breve{\mathcal{A}}$ complex spin-0 field, etc.

In the case of a $p=2$ para-Family,  the generalized ray representatives also occur in ray multiplets.  However, because of these ``flip" reordering relations,  there are four ``Kinds" of ray multiplets due to the occurrence of four Kinds of possible ``small centers" in the associated bracket operator ray representative.  

The subscript $K$ is used to denote which Kind.  The generalized ray representative is labeled by
\ber
 | r, \rho > |_{_{K}} ={Y}^{ l(r) } _n {\mathcal Y}^{ \lambda(\rho) } _\nu |_{_{K,s}}
\eer
This ray occurs in the generally truncated ray multiplet 
\ber
\overline{Y}^{ l_{min} (s) } _n   \overline{\mathcal Y}^{ \lambda_{min}(s)} _\nu |_{_{K}}= \overline{Y} _n   \overline{\mathcal Y} _\nu |_{_{K,s}}
\eer
For specifying the para-Family ray and ray multiplet, there is an ``$s$  parameter" which both determines the length of the ``small center," see eqs. (42-43), and determines the values of the two ``truncation functions" $ l_{min} (s)$ and $ \lambda_{min}(s)$ for the ray multiplet.  

Like the ``$r$" and ``$\rho$" integer parameters for the pB and pF ray multiplets, this additional ``$s$ parameter"  is integer valued with a range $ s=0, 1, \cdots s_{max}$ where $s_{max} \approx \frac{n_T}{2}$ depends on which Kind of ray multiplet. It is bounded $s_{max} \leq  \frac{n_T}{2} $.   

These para-Family ray multiplets can be collected in one of four Kinds of more inclusive ``Meta-Multiplets" $\overline{\overline{Y} }_n   \overline{\overline{\mathcal Y}} _\nu |_{_{K}}$ with $s$ labeling its successive truncated ray multiplets $ \overline{Y}_n   \overline{\mathcal Y} _\nu |_{_{K,s}} $
\ber
\overline{\overline{Y} }_n  \overline{ \overline{\mathcal Y}} _\nu |_{_{K}} = \{ \overline{Y}_n   \overline{\mathcal Y} _\nu |_{_{K,s=0}}, \cdots,\overline{Y}_n   \overline{\mathcal Y} _\nu |_{_{K,s_{max}}}        \}    
\eer
This is analogous to the use of $r$ and $\rho$ for the labeling  of pB-rays by ${Y}^{ l(r) } _n$ and of  pF rays by  ${\mathcal Y}^{ \lambda(\rho) } _\nu$ in their respective ray multiplets $\overline{Y} _n$ and $ \overline{\mathcal Y} _\nu$.   Note:  $s_{min} = 0 $ whereas $r_{min} = 1 $ and ${\rho}_{min} = 1 $.

Given $( n, \nu, K, s)$ the pFam generalized ray multiplet $ \overline{Y}_n   \overline{\mathcal Y} _\nu |_{_{K,s}} $ is uniquely determined.

Caution:  The Young-diagram parts of the notational labeling of the irreducible rays and ray multiplets is the same for both the para-Family case $ \overline{Y}_n  ~  \overline{\mathcal Y}_{\nu} $ and the direct product case $ \overline{Y}_n \times  \overline{\mathcal Y}_{\nu} $.  However, due the trilinear relations for $ \overline{Y}_n  ~  \overline{\mathcal Y}_{\nu} $ case,  the algebraic operator properties of the generalized ray representatives are obviously very different for the two cases.    This operator distinction must also be remembered in considering different rays (labeled by the same Young-diagram parts of the notation) in different truncated ray multiplets $ \overline{Y} _n   \bar{\mathcal Y} _\nu |_{_{K,s}} $ in the same ``Meta-Multiplet" $\overline{\overline{Y} }_n   \overline{\overline{\mathcal Y}} _\nu |_{_{K}}$.

\subsubsection{\label{sec:level2} Four Kinds of Small Centers \newline in Occupation Number Basis}

The expressions below are to be read Left-to-Right.  The vacuum state on the extreme right is omitted in displaying these state expressions:

For this $n$-paraboson and $\nu$-fermion system, we also introduce $s$-dependent $\bar{n}(s)$ and $\bar{\nu}(s)$ non-negative integer-valued functions which differ with the four Kind's of small centers.    These two barred functions are used to count the number of pB (pF) creation operators to the left (right) of the small centers. 
  
In the occupation number basis, the K1 ray has the form:
\ber
(a^{\dag} b^{\dag} \cdots e^{\dag} f^{\dag}) (\phi ^{\dag} w^{\dag} \chi^{\dag} x^{\dag} \cdots  \psi^{\dag} y^{\dag}  \omega^{\dag} z^{\dag}) ( \delta^{\dag} \gamma^{\dag} \cdots \beta^{\dag} \alpha^{\dag}) \nonumber \\
\eer
For each such ``ONB-ray" (Occupation Number Basis ray), irrespective of its ``Kind," the left factor always has an even-number $ 2 \bar{n}(s)$ pB creation operators, and the right factor always has an even-number $ 2 \bar{\nu}(s)$ pF creation operators.  

The middle factor, named the ``small center," alternates in pB  and pF  operators.  For the K1 ray
\ber
c(s) |_{_{K1}}= (\phi ^{\dag} w^{\dag} \chi^{\dag} x^{\dag} \cdots  \psi^{\dag} y^{\dag}  \omega^{\dag} z^{\dag})  
\eer
The K1 small center starts with a pF creation operator and has $2s$ operators.

The small center always starts with an operator in the ``odd" place position. 
These small center factors are totally symmetric in pB operators and totally antisymmetric in pF operators.  

 Versus other small centers which always contain some creation operators, for $s_1=0$ the K1 small center is special in that it is the identity operator.   It is sometimes useful to put a ``Kind" subscript on the $s$-parameter.
The total number of creation operators, $n_T = n + \nu$, in the K1 ray (42) is
\ber
n_T = 2 \bar{n}(s) + 2 \bar{\nu}(s) + 2 s   \nonumber
\eer

This K1 ONB-ray is used below to construct the bracket ray representatives for the
\ber
\overline{Y}^{ s } _n  ~  \overline{\mathcal Y}^{ s} _\nu = \overline{Y}^{ s } _{2 \bar{n} +s}   \overline{\mathcal Y}^{ s} _{2 \bar{\nu} +s} 
\eer
ray multiplet.   For K1, the four $s$-dependent equations are $n = 2 \bar{n}(s) +s$, $\nu = 2 \bar{\nu}(s) +s$, $l_{min}(s) =s$, $\lambda_{min}(s) =s$, with $s_{max}=  \frac{n_T}{2}$.

The K2 ONB-ray also has a total even number of operators in it.   So for K1 and K2, either $n$ and $\nu$ are both even-valued or both are odd-valued.   Unlike the K1 small center in (42) and (43), the K2 small center starts with a pB creation operator and has $2s + 2$ operators.   The K2 ray has the form:
\ber
(a^{\dag} b^{\dag} \cdots e^{\dag} f^{\dag}) (w^{\dag} \phi ^{\dag} x^{\dag}  \chi^{\dag}  \cdots y^{\dag} \psi^{\dag} z^{\dag} \omega^{\dag}) ( \delta^{\dag} \gamma^{\dag} \cdots \beta^{\dag} \alpha^{\dag}) \nonumber \\
\eer
The total number of creation operators in the K2 ray is
\ber
n_T = 2 \bar{n}(s) + 2 \bar{\nu}(s) + 2 s + 2  \nonumber
\eer

This K2 ONB-ray is used below to construct the bracket ray representatives for the
\ber
\overline{Y}^{ s+1 } _n  ~  \overline{\mathcal Y}^{ s+1} _\nu = \overline{Y}^{ s+1 } _{2 \bar{n} +s+1}   \overline{\mathcal Y}^{ s+1} _{2 \bar{\nu} +s+1} 
\eer
ray multiplet.   For K2, $n = 2 \bar{n}(s) +s+1$, $\nu = 2 \bar{\nu}(s) +s+1$, $l_{min}(s) =s+1$, $\lambda_{min}(s) =s+1$, with $s_{max} =  \frac{n_T}{2}- 1$.

In $p=2$ parastatistics, place-positions are not directly physical [8, 5] but are often very important, such as here in classifying small centers and then obtaining the associated pFam ray representatives.  For example, the Young Diagram labeling is the same for the above two small centers, one for K1 and the other for K2,  with the choice of their respective $s$-parameters $s_1 = s_2 +1 \neq 0$.  These small centers differ only in the place-position ordering of the pB and pF operators. In the simplest case of $s_1 =1$, there is $ \phi ^{\dag} w^{\dag}$ for K1 and $w^{\dag} \phi ^{\dag}$ for K2.    These operators are the same in the case of a direct product $ \overline{Y} \times  \overline{\mathcal Y} $ system.  By the assumed absence of ``second unit observables" (see Appendix B in [5]), in a pFam system $ \overline{Y}_n  ~  \overline{\mathcal Y}_{\nu} $ it is the superposition $ \frac{1}{\sqrt{2}}  \{\phi ^{\dag} , w^{\dag}\}$ which is coupled to a physical observable.  The orthogonal $ \frac{1}{\sqrt{2}} [ \phi ^{\dag} , w^{\dag}  ]$ is not coupled.  

There are two physical initial states in the case of a pB and pF collision for a pFam system $ \overline{Y}  ~  \overline{\mathcal Y} $ but only one initial state for the case of a direct product $ \overline{Y} \times  \overline{\mathcal Y} $ system.   By completeness of states for the initial states, in this counting of states for a pFam system, there are two states whether the orthonormal basis used is $ \phi ^{\dag} w^{\dag}$ and $w^{\dag} \phi ^{\dag}$, or instead a basis with the orthonormal superpositions $ \frac{1}{\sqrt{2}}  \{\phi ^{\dag} , w^{\dag}\}$ and $ \frac{1}{\sqrt{2}} [ \phi ^{\dag} , w^{\dag}  ]$.

The two remaining ONB-rays K3 and K4 have a total odd number of operators in them, so either $n$ or $\nu$ is odd-valued with the other even-valued:   The K3 small center starts with a pF creation operator and has $2s + 1$ operators.   It has the form:
\ber
(a^{\dag} b^{\dag} \cdots e^{\dag} f^{\dag}) (\phi ^{\dag} w^{\dag} \chi^{\dag} x^{\dag} \cdots y^{\dag} \psi^{\dag} z^{\dag} \omega^{\dag} ) ( \delta^{\dag} \gamma^{\dag} \cdots \beta^{\dag} \alpha^{\dag}) \nonumber \\
\eer
The total number of creation operators in the K3 ray is
\ber
n_T = 2 \bar{n}(s) + 2 \bar{\nu}(s) + 2 s + 1  \nonumber
\eer

This K3 ONB-ray is used below to construct the bracket ray representatives for the
\ber
\overline{Y}^{ s } _n  ~  \overline{\mathcal Y}^{ s+1} _\nu = \overline{Y}^{ s } _{2 \bar{n} +s}   \overline{\mathcal Y}^{ s+1} _{2 \bar{\nu} +s+1} 
\eer
ray multiplet.   For K3, $n = 2 \bar{n}(s) +s$, $\nu = 2 \bar{\nu}(s) +s+1$, $l_{min}(s) =s$, $\lambda_{min}(s) =s+1$, with $s_{max} =  \frac{n_T - 1}{2}$.

The K4 small center starts with a pB operator.  As for K3, K4 has $2s + 1$ operators.  The K4 ONB-ray has the form:
\ber
(a^{\dag} b^{\dag} \cdots e^{\dag} f^{\dag}) (w^{\dag} \phi ^{\dag} x^{\dag}  \chi^{\dag}  \cdots \psi^{\dag} y^{\dag}  \omega^{\dag} z^{\dag}) ( \delta^{\dag} \gamma^{\dag} \cdots \beta^{\dag} \alpha^{\dag}) \nonumber \\
\eer
The total number of creation operators in the K4 ray is
\ber
n_T = 2 \bar{n}(s) + 2 \bar{\nu}(s) + 2 s + 1  \nonumber
\eer

This K4 ONB-ray is used below to construct the bracket ray representatives for the
\ber
\overline{Y}^{ s+1 } _n  ~  \overline{\mathcal Y}^{ s} _\nu = \overline{Y}^{ s+1 } _{2 \bar{n} +s+1}   \overline{\mathcal Y}^{ s} _{2 \bar{\nu} +s} 
\eer
ray multiplet.   For K4, $n = 2 \bar{n}(s) +s+1$, $\nu = 2 \bar{\nu}(s) +s$, $l_{min}(s) =s+1$, $\lambda_{min}(s) =s$, with $s_{max} =  \frac{n_T - 1}{2}$. 

In the Young diagram parts of the labeling, K4 and K3 are related by interchanging the pB and pF labels.

These para-Family ONB-rays are complete.  In particular, the ONB-rays obtained by writing instead a pF factor on the left of the small center with a pB factor on the right are equivatlent to the ONB-rays listed above.

When $\bar{n}(s)=\bar{\nu}(s)=0$, in the ONB-rays there is only the small center factor.  The associated Young-diagram labeling of the rays is the same as for the 1-dimensional ones used for ordinary bosons and fermions.  But for $n_T$ even valued, for the pFam there is the $s_1=s_2 +1 \neq 0$ place-position duplication for K1 and K2 rays. 

\subsubsection{\label{sec:level2} Four Generic Functions Which are Independent of Kind }

The generalized ray representative 
\newline
$ ~~~~~~~~~~~  | r, \rho > |_{_{K}} =
{Y}^{ l(r) } _n {\mathcal Y}^{ \lambda(\rho) } _\nu |_{_{K,s}}
$
\newline
in the ray multiplet 
\newline
$~~~~~~~~~~~ \overline{Y}^{ l_{min} (s) } _n   \overline{\mathcal Y}^{ \lambda_{min}(s)} _\nu |_{_{K}}$
\newline
is labeled by the two Kind-generic functions
\ber
l(r)=n-2(r-1) \nonumber \\
\lambda(\rho)= \nu - 2 (\rho -1).   
\eer
which determine the length of the legs of the pB and pF Young Diagrams which partially label the rays.  

Analogous to the case of only pB's, in the ray multiplet $ \overline{Y}_n  ~  \overline{\mathcal Y}_{\nu} $ this ray $| r, \rho > $ is labeled by the value of  $r=1, \cdots r_{max}(s)$ with $r_{max}(s)=\bar{n}(s) +1$ and the value of  $\rho=1, \cdots \rho_{max}(s)$ with $\rho_{max}(s)=\bar{\nu}(s) +1$.  

Conversely, given the pB Young Diagram $n$ and $l$ values, $r=\frac{1}{2} (n-l+2)$, and given the pF Young Diagram $\nu$ and $\lambda$ values, $\rho=\frac{1}{2} (\nu-\lambda+2)$.
The four $s$-dependent functions $ l_{min} (s),\lambda_{min}(s), \bar{n}(s), \bar{\nu}(s) $  depend on the Kind of para-Family.  These expressions are listed in the text in the preceding subsection on the four Kinds of small centers, respectively following (44, 46, 48, 50).

The total number of rays in the rays in the ray multiplet $\overline{Y} _n   \overline{\mathcal Y} _\nu |_{_{K,s}}$ is the $s$-dependent product 
\ber
r_{max}(s)  ~ \rho_{max}(s)      = [ \bar{n}(s) +1  ]  [\bar{\nu}(s)   +1    ]         \nonumber 
\eer
The expressions for $r_{max}(s)$ and $\rho_{max}(s)$ depend on the Kind of ray multiplet and are given in the next subsection with the pFam generalized ray representatives  (56, 59, 62, 65).

The bracket constructions listed below will also use the two Kind-generic even-valued, non-negative functions 
\ber
h(r)_{max}=n- l(r) = 2r-2 ,    \nonumber \\
\iota(\rho)_{max}= \nu-\lambda(\rho)= 2 \rho -2  
\eer
These two functions respectively denote the total Even-number of pB operators in the commutator factors on the left $( \frac{1}{\sqrt{2}} [ a^{\dag}_{h-1}, a^{\dag}_h] )$, and the total Even-number of pF operators in the anti-commutator factors on the right $( \frac{1}{\sqrt{2}} \{ {\alpha}^{\dag}_{{\iota}}, {\alpha}^{\dag}_{\iota-1} \}) $ in the bracket construction for the generalized ray representative for a $p=2$ para-Family, see 53) immediately below.

\subsubsection{\label{sec:level2} Bracket Generalized Ray Representatives for a Para-Family }

For the ``K1 Kind" of ray multiplet with an even number of paraparticles
\ber
\overline{Y}^{ s } _n  ~  \overline{\mathcal Y}^{ s} _\nu = \overline{Y}^{ s } _{2 \bar{n} +s}   \overline{\mathcal Y}^{ s} _{2 \bar{\nu} +s}, \nonumber 
\eer
the generalized ray representative is
\ber
 | r, \rho > |_{_{K1}} = \prod^{h_{max}(r)}_{h=2 } ( \frac{1}{\sqrt{2}} [ a^{\dag}_{h-1}, a^{\dag}_h] ) ~~ \bar {C}_{_{K1}}(s,l(r),\lambda(\rho)) \nonumber \\
 \prod^{{\iota}_{max}(\rho)}_{{\iota}=2 } ( \frac{1}{\sqrt{2}} \{ {\alpha}^{\dag}_{{\iota}}, {\alpha}^{\dag}_{\iota-1} \}) ~~~~~
\eer
with the ``Full-Center" for K1 
\ber
 \bar {C}_{_{K1}}(s,l(r),\lambda(\rho))=   \frac{1}{\sqrt{N_1(l, s) \mathcal{N}_1(\lambda, s)}}   \nonumber \\
\big\{     \sum\limits_{\pi \varepsilon { S}_l }          \sum\limits_{\sigma \varepsilon { \mathcal S}_\lambda }  (-)^{\sigma}   b_{\pi (h_{max} +1)}^{\dag} \cdots   b_{\pi (n-s)}^{\dag}  \nonumber \\
   \big(  \beta_{\sigma (\nu )}^{\dag}  b_{\pi (n-s+1)}^{\dag}  \cdots        \beta_{\sigma (\nu - s+1}^{\dag}  b_{\pi (n)}^{\dag}   \big)  \nonumber \\
   \beta_{\sigma (\nu - s)}^{\dag}  \cdots \beta_{\sigma (\iota_{max} +1)}^{\dag} \big\}                  
\eer
Note in the Full-Center $\bar {C}_{_{K1}}$ the small center expression $c(s) |_{_{K1}}$ of (43) occurs.

The format of this sandwich-like operator formula (53) is also ``generic."  The only dependence on the Kind of the para-Family ray is in the Full-Center factor.   Below, for each of the other 3 Kinds, only the Full-Center factor is listed.  

In this  $\bar {C}_{_{K1}}$ expression, between the curly braces there is first a full permutation of the pB operators by the ${\pi \varepsilon { S}_l }$ to achieve a distinct set of terms, and an analogous full permutation of the pF operators by the ${\sigma \varepsilon { \mathcal S}_\lambda } $.  The $(-)^{\sigma}$ factor gives a negative sign for an odd number of pF-exchanges  $ \overline{\mathcal P}_{\iota,\kappa}$.  Next there is use of the ``flip" SCR relations to minimize the number of terms to $N_1(l, s) \mathcal{N}_1(\lambda, s)$.  This normalization factor equals the number of independent terms in $\bar {C}_{_{K1}}$. It is the product of 
\ber
N_1(l, s) = \frac{l !}{(\frac{l+s}{2}) !(\frac{l-s}{2} )!} \nonumber \\
\mathcal{N}_1(\lambda, s) = \frac{\lambda !}{(\frac{\lambda+s}{2} )! (\frac{\lambda-s}{2} )!}
\eer
This minimization procedure for $p=2$ parastatistics, to obtain the minimal independent terms, is the direct generalization of the pB/pF ``unprimed" bracket procedure of Section II for the cases of the simpler pB/pF generalized ray spaces, see (5, 6).

In Full-Center $\bar {C}_{_{K}}$, the number of pB creation operators on the left is always $2(r_{max}(s))-r)$.  The number of pF creation operators on the right is always $ 2(\rho_{max}(s)-\rho)$.  

For $\bar {C}_{_{K1}}$, there are 
\ber
2(r_{max}(s)-r) = l(r)-s   \nonumber 
\eer
pB operators on the left, and there are
\ber
2(\rho_{max}(s)-\rho) = \lambda(\rho) -s  \nonumber 
\eer
pF operators on the right.
Also unique to the K1 ray multiplet are the formulas:
\ber
r_{max}(s)= \frac{1}{2}(n-s+2), \rho_{max}(s)=\frac{1}{2}(\nu-s+2), \nonumber \\
\eer
Their product gives the total number of rays in the K1 ray multiplet.  

The K1 ``Meta-Multiplet" is 
\ber
\overline{\overline{Y} }_n  \overline{ \overline{\mathcal Y}} _\nu |_{_{K1}} = \{ \overline{Y}^{0}_n   \overline{\mathcal Y}^{0} _\nu, \cdots, \overline{Y}^{s}_n   \overline{\mathcal Y}^{s} _\nu,\cdots,\overline{Y}^{ s_{max}}_n   \overline{\mathcal Y}^{ s_{max}} _\nu        \}   |_{_{K1}}  \nonumber
\eer
with $ s=0, 1, \cdots s_{max}$ with $s_{max} =  \frac{n_T}{2}$.

For the ``K2 Kind" of ray multiplet with also an even number of paraparticles
\ber
\overline{Y}^{ s+1 } _n  ~  \overline{\mathcal Y}^{ s+1} _\nu = \overline{Y}^{ s +1} _{2 \bar{n} +s+1}   \overline{\mathcal Y}^{ s+1} _{2 \bar{\nu} +s+1}, \nonumber 
\eer
the generalized ray representative bracket expression has the same form as in (53) but
with the Full-Center for K2 
\ber
 \bar {C}_{_{K2}}(s,l(r),\lambda(\rho))=   \frac{1}{\sqrt{N_2(l, s) \mathcal{N}_2(\lambda, s)}}  \nonumber \\
 \big\{     \sum\limits_{\pi \varepsilon { S}_l }          \sum\limits_{\sigma \varepsilon { \mathcal S}_\lambda }           (-)^{\sigma}   b_{\pi (h_{max} +1)}^{\dag} \cdots  b_{\pi (n-s-1)}^{\dag} \nonumber \\
  \big(   b_{\pi (n-s)}^{\dag}  \beta_{\sigma (\nu )}^{\dag} \cdots         b_{\pi (n)}^{\dag}  \beta_{\sigma (\nu - s}^{\dag}  \big)  \nonumber \\
  \beta_{\sigma (\nu - s-1)}^{\dag}  \cdots \beta_{\sigma (\iota_{max} +1)}^{\dag} \big\}                  
\eer
with the small center of (45).

For this Full-Center  $\bar {C}_{_{K2}}$, 
\ber
N_2(l, s) = \frac{l !}{(\frac{l+s+1}{2}) !(\frac{l-s-1}{2} )!} \nonumber \\
\mathcal{N}_2(\lambda, s) = \frac{\lambda !}{(\frac{\lambda+s+1}{2} )! (\frac{\lambda-s-1}{2} )!}
\eer
there are 
\ber
2(r_{max}(s)-r) = l(r)-s-1   \nonumber 
\eer
pB operators on the left, and there are
\ber
2(\rho_{max}(s)-\rho) = \lambda(\rho) -s -1 \nonumber 
\eer
pF operators on the right.
Unique to the K2 ray multiplet are the formulas:
\ber
r_{max}(s) = \frac{1}{2}(n-s+1), \rho_{max}(s)=\frac{1}{2}(\nu-s+1), \nonumber \\
\eer

The K2 ``Meta-Multiplet" is 
\ber
\overline{\overline{Y} }_n  \overline{ \overline{\mathcal Y}} _\nu |_{_{K2}} = \{ \overline{Y}^{1}_n   \overline{\mathcal Y}^{1} _\nu, \cdots,  \overline{Y}^{s+1}_n   \overline{\mathcal Y}^{s+1} _\nu, \nonumber \\
\cdots,\overline{Y}^{ s_{max+1}}_n   \overline{\mathcal Y}^{ s_{max+1}} _\nu        \}   |_{_{K2}}  \nonumber
\eer
with $ s=0, 1, \cdots s_{max}$ with $s_{max} =  \frac{n_T}{2}-1$.

The ``K3 Kind" of ray multiplet has an odd number of paraparticles
\ber
\overline{Y}^{ s } _n  ~  \overline{\mathcal Y}^{ s+1} _\nu = \overline{Y}^{ s } _{2 \bar{n} +s}   \overline{\mathcal Y}^{ s+1} _{2 \bar{\nu} +s+1}, \nonumber 
\eer
The generalized ray representative bracket expression has the same form as in (53) but
with the Full-Center for K3 
\ber
 \bar {C}_{_{K3}}(s,l(r),\lambda(\rho))=    \frac{1}{\sqrt{N_3(l, s) \mathcal{N}_3(\lambda, s)}}  \nonumber \\
 \big\{     \sum\limits_{\pi \varepsilon { S}_l }          \sum\limits_{\sigma \varepsilon { \mathcal S}_\lambda }           (-)^{\sigma}   b_{\pi (h_{max} +1)}^{\dag} \cdots  b_{\pi (n-s)}^{\dag} \nonumber \\
  \big(  \beta_{\sigma (\nu )}^{\dag}  b_{\pi (n-s+1)}^{\dag}  \cdots     b_{\pi (n)}^{\dag}      \beta_{\sigma (\nu - s}^{\dag}   \big)   \nonumber \\
  \beta_{\sigma (\nu - s-1)}^{\dag}  \cdots \beta_{\sigma (\iota_{max} +1)}^{\dag} \big\}                 
\eer
with the small center of (47).

For this Full-Center  $\bar {C}_{_{K3}}$, 
\ber
N_3(l, s) = \frac{l !}{(\frac{l+s}{2}) !(\frac{l-s}{2} )!} \nonumber \\
\mathcal{N}_3(\lambda, s) = \frac{\lambda !}{(\frac{\lambda+s+1}{2} )! (\frac{\lambda-s-1}{2} )!}
\eer
There are 
\ber
2(r_{max}(s)-r) = l(r)-s   \nonumber 
\eer
pB operators on the left, and there are
\ber
2(\rho_{max}(s)-\rho) = \lambda(\rho) -s -1 \nonumber 
\eer
pF operators on the right.
Unique to the K3 ray multiplet are the formulas:
\ber
r_{max}(s) = \frac{1}{2}(n-s+2), \rho_{max}(s)=\frac{1}{2}(\nu-s+1), \nonumber \\
\eer

The K3 ``Meta-Multiplet" is 
\ber
\overline{\overline{Y} }_n  \overline{ \overline{\mathcal Y}} _\nu |_{_{K3}} = \{ \overline{Y}^{0}_n   \overline{\mathcal Y}^{1} _\nu, \cdots, \overline{Y}^{s}_n   \overline{\mathcal Y}^{s+1} _\nu,  \nonumber \\
\cdots, \overline{Y}^{ s_{max}}_n   \overline{\mathcal Y}^{ s_{max}+1} _\nu        \}   |_{_{K3}}  \nonumber
\eer
with $ s=0, 1, \cdots s_{max}$ with $s_{max} =  \frac{n_T - 1}{2}$.

The ``K4 Kind" of ray multiplet also has an odd number of paraparticles
\ber
\overline{Y}^{ s+1 } _n  ~  \overline{\mathcal Y}^{ s} _\nu = \overline{Y}^{ s+1 } _{2 \bar{n} +s+1}   \overline{\mathcal Y}^{ s} _{2 \bar{\nu} +s}, \nonumber 
\eer
The generalized ray representative bracket expression has the same form as in (53) but
with the Full-Center for K4 
\ber
 \bar {C}_{_{K4}}(s,l(r),\lambda(\rho))=   \frac{1}{\sqrt{N_4(l, s) \mathcal{N}_4(\lambda, s)}}  \nonumber \\
  \big\{     \sum\limits_{\pi \varepsilon { S}_l }          \sum\limits_{\sigma \varepsilon { \mathcal S}_\lambda }         (-)^{\sigma}   b_{\pi (h_{max} +1)}^{\dag} \cdots  b_{\pi (n-s-1)}^{\dag} \nonumber \\
  \big(   b_{\pi (n-s)}^{\dag}  \beta_{\sigma (\nu )}^{\dag} \cdots           \beta_{\sigma (\nu - s+1}^{\dag} b_{\pi (n)}^{\dag}  \big)  \nonumber \\
  \beta_{\sigma (\nu - s)}^{\dag}  \cdots \beta_{\sigma (\iota_{max} +1)}^{\dag} \big\}              
\eer
with the small center of (49).

For this Full-Center  $\bar {C}_{_{K4}}$, 
\ber
N_4(l, s) = \frac{l !}{(\frac{l+s+1}{2}) !(\frac{l-s-1}{2} )!} \nonumber \\
\mathcal{N}_4(\lambda, s) = \frac{\lambda !}{(\frac{\lambda+s}{2} )! (\frac{\lambda-s}{2} )!}
\eer
there are 
\ber
2(r_{max}(s)-r) = l(r)-s-1   \nonumber 
\eer
pB operators on the left, and there are
\ber
2(\rho_{max}(s)-\rho) = \lambda(\rho) -s  \nonumber 
\eer
pF operators on the right.
Unique to the K4 ray multiplet are the formulas:
\ber
r_{max}(s) = \frac{1}{2}(n-s+1), \rho_{max}(s)=\frac{1}{2}(\nu-s+2), \nonumber \\
\eer

The K4 ``Meta-Multiplet" is 
\ber
\overline{\overline{Y} }_n  \overline{ \overline{\mathcal Y}} _\nu |_{_{K4}} = \{ \overline{Y}^{1}_n   \overline{\mathcal Y}^{0} _\nu, \cdots, \overline{Y}^{s+1}_n   \overline{\mathcal Y}^{s} _\nu,   \nonumber \\
\cdots,\overline{Y}^{ s_{max}+1}_n   \overline{\mathcal Y}^{ s_{max}} _\nu        \}   |_{_{K4}}  \nonumber
\eer
with $ s=0, 1, \cdots s_{max}$ with $s_{max} =  \frac{n_T-1}{2}$.

If a phase convention for generalized rays in $p=2$ parastatistics is required, a simple one for each Kind of  $| r, \rho > |_{_{K}} $ ray with the format of (53) is to assign a plus overall sign for a ray with the first term with its $i, j, \cdots$ pB operators labeled from the left and its $\iota, \kappa, \cdots$ pF operators labeled from the right.   In the absence of pF's (of pB's) this convention also applies with $s_1=0$ for $n_T=n+\nu$ even valued, and $s_3=0$ ($s_4=0$) for $\nu$ ($n$) odd-valued.  For a direct product $ \overline{Y}_n \times  \overline{\mathcal Y}_{\nu} $ system, this convention is for the pF operators.
 
 \section{Generalized Ray Spaces for Arbitrary $p$ Order}
 
 As the $p$ value increases, more generalized rays occur for a fixed number of parabosons so the dimension $\overline{d}_{n}$ of the ray multiplet $ \overline Y^{ } _n $ increases, $ \overline {d}_{n} =$ (sum of dimensions of rays in $ \overline Y^{ } _n $).  
 
 Both the dimension $d_{n}$ of the ray subspace $ Y^{ l } _{n}$ and the generalized ray's $  {\overline P} _{sum} $ eigenvalue $e$ are independent of the $p$ value.  The  non-negative $n_{s,a }$ respectively monotonically decrease (increase) as $r$ increases through the paraboson ray multiplet.
 For rays with Young Diagrams which occur for $p=2$, the full labeling of a ray subspace $ \prescript{}{d} Y^{ l } _{n,e}$ is independent of the $p$ value.   These $p$ dependence/independence properties analogously hold for parafermions and para-Families.   
 
 For $p \geq 3$ and  pB Young Diagrams with $r\geq 3$ rows, the $l$ label for specifying the generalized ray for $p=2$ can be expanded to the set $\{ l_1, \cdots l_{r-1}  \}$ where $l_i$=(extra length of preceding row), so there is a distinct minimal labeling
$ Y^{ \{ l_1, \cdots l_{r-1}  \} } _{n}$. 

For calculation of the normalization factors for the following ray representatives for arbitrary order $p$, the representation of the explicit ray with Green indices was used, see Appendix B.   Any matrix element of any operator expressed as a product of paraparticle creation and annihilation operators, with free paraparticle initial and final states, can be evaluated using Green indices.
The norm for any state can thereby be obtained. In parastatistics, up to $p$ identical pB's (pF's) can occupy a totally antisymmetric (symmetric) state.   The norm of these one-dimensional states/rays is proportional to $p!$.
 
\subsection{\label{sec:level2}Explicit orthonormal ray representatives for a few paraparticles }

For arbitrary $p$ order, for $2$ parabosons corresponding to the two 1-dimensional irreducible representations  $ Y^{ 2 } _2=\prescript{}{1} Y^{ 2} _{2,1}$, $Y^{ 0} _2=\prescript{}{1} Y^{ 0 } _{2,-1}$ the orthonormal rays are
\ber
\frac{1}{  2 \sqrt{p}  }    \{  {a}^{\dag} ,{b}^{\dag}  \}       |0> ,
\frac{1}{2 \sqrt{p(p-1)} }  [{a}^{\dag} ,{b}^{\dag} ] |0>
\eer

The $p$-dependent normalization factor is singular when the state cannot be normalized because the $p$ value is too small, so the state does not exist.  This occurs in (66) for the ordinary boson value $p=1$.  

For $2$  parafermions, $ {\mathcal Y}^{ l } _n = \mathcal Y^{ 2 } _2=\prescript{}{1} {\mathcal Y}^{ 2} _{2,-1} $ and $ {\mathcal Y}^{ 0} _2=\prescript{}{1} {\mathcal Y}^{ 0} _{2,1}  $  the rays  are
\ber
\frac{1}{2 \sqrt{p}}      [  {\alpha}^{\dag} ,{\beta}^{\dag}  ]       |0> ,
\frac{1}{2 \sqrt{p(p-1)}} \{{\alpha}^{\dag} ,{\beta}^{\dag} \} |0>
\eer

For the commuting pB times pF case  
$ {Y}^{ 1} _1 \times {\mathcal Y}^{1} _1 $, the 2 particle ray representative is
\ber
\frac{1}{  2 \sqrt{p}  }    \{  {a}^{\dag} ,{\beta}^{\dag}  \}       |0> ,
\eer

For the para-Family case, the 2 particle rays are
\ber
\frac{1}{  2 \sqrt{p}  }    \{  {a}^{\dag} ,{\beta}^{\dag}  \}       |0> ,
\frac{1}{2 \sqrt{p(p-1)} }  [{a}^{\dag} ,{\beta}^{\dag} ] |0>
\eer

For $3$ parabosons, for $p \geq 3$ the ray multiplet $   \overline Y^{} _3 $ has three rays  $ Y^{ 3 } _3 = \prescript{}{1} Y^{ 3} _{3,3}$, $  Y^{ 1 } _3 = \prescript{}{2} Y^{ 1 } _{3,0}$, $ Y^{ 0 } _3 = \prescript{}{1} Y^{ 0} _{3,-3}$ 
\newline
with respective ray representatives
\ber
\frac{1}{2\sqrt{3p(p+2)}}      \{  {a}^{\dag} ,{b}^{\dag} , {c}^{\dag} \} ^{'}      |0> ,   \nonumber \\
\frac{1}{4\sqrt{2p(p-1)}}[ [{a}^{\dag} ,{b}^{\dag}] ,{c}^{\dag}]|0>, \nonumber \\
\frac{1}{6 \sqrt{p(p-1)(p-2)}}      [  {a}^{\dag} ,{b}^{\dag} , {c}^{\dag} ] ^{'}      |0>
\eer
In the first and last rays, a ``prime" superscript has been added to the totally symmetric (totally antisymmetric) brackets to indicate that in these brackets the ``flip" reordering relations special to orders $p=1,2$ have not been used; see discussion following (5) above.  The second ray representative has an outside commutator bracket versus the formula for $p=2$ in (7).

For $3$ parafermions, the ray multiplet $   \overline {\mathcal Y}^{} _3 $ for $p \geq 3$ has the three rays  $ {\mathcal Y}^{ 3 } _3 = \prescript{}{1} {\mathcal Y}^{ 3} _{3,-3}$, $  {\mathcal Y}^{ 1 } _3 = \prescript{}{2} {\mathcal Y}^{ 1 } _{3,0}$, $ {\mathcal Y}^{ 0 } _3 = \prescript{}{1} {\mathcal Y}^{ 0} _{3,3}$ 
\ber
\frac{1}{2\sqrt{3p(p+2)}}     [  {\alpha}^{\dag} ,{\beta}^{\dag} , {\gamma}^{\dag} ] ^{'}      |0> ,   \nonumber \\
\frac{1}{4\sqrt{2p(p-1)}}[ \{  {\alpha}^{\dag} ,{\beta}^{\dag}\},{\gamma}^{\dag}]|0>,   \nonumber    \\
\frac{1}{6 \sqrt{p(p-1)(p-2)}}    \{  {\alpha}^{\dag} ,{\beta}^{\dag} , {\gamma}^{\dag} \}^{'}      |0>
\eer
The second ray representative has an outside commutator bracket versus the formula in (28) for $p=2$.

\section{Exchanging Generalized Ray Operators }

Exchange phenomena in quantum mechanical scattering processes are a very important application of identical particle statistics.   Exchange properties are very useful in symmetry tests, in statistical mechanics, and in condensed matter systems.   One use of the ``exchange relations" considered in this section would be in the analysis of a decay or scattering process with two final rays in different regions of coordinate space.    Neutral dark matter paraparticles might occur as in the ``statistics portal"  Wess-Zumino model with spin zero particle $\breve A$ and antiparticle  $\breve{B}= \overline{\breve{A}}$  plus a spin 1/2 Majorana   $\breve{\nu}$, see [5].    Depending on whether $p = 2$ or $p \geq 3$, the two generalized ray operators often commute or don't commute, see below.   

For generalized rays with a few paraparticles, we systematically list the nested brackets of totally symmetric and totally antisymmetric creation operators for pB's, for pF's, and for pFam's.   For $p=1$,  $ [ {a}^{\dag} ,{b}^{\dag} ]=0$ and $\{ {\alpha}^{\dag} ,{\beta}^{\dag} \}=0$ which are awkward caveats to frequently note, so in this Section we often refer to ``$p=2$, $p \geq 2$ and ``$p \geq 3$," instead of to ``arbitrary $p$."  Most of the results listed here follow by using Green indices, see Appendix B.   Once a nested bracket is non-zero for $p=3$ it must be for $p\geq 3$.  

\subsection{\label{sec:level2}Paraboson Ray Exchanges }

For 1 and 2 particle ray exchanges:   For $p \geq 2$ 
\ber
[{a}^{\dag}, \{ {b}^{\dag} ,{c}^{\dag} \} ] =0, \{ {a}^{\dag}, \{ {b}^{\dag} ,{c}^{\dag}\} \}\neq 0,\nonumber 
\eer
\ber
 [ {a}^{\dag}, [ {b}^{\dag} ,{c}^{\dag} ] ]\neq 0
\eer
For $p = 2$, 
\ber
\{ {a}^{\dag}, [ {b}^{\dag} ,{c}^{\dag} ] \} = 0
\eer
but not zero for $p \geq 3$.  Hence, for $p\geq 3$ there is a ``non-commutative exchange" for the rays $ Y^{1}_1$ and $ Y^{0}_2$.   $ Y^{0}_2$ is the 2 paraboson antisymmetric ray which does not occur for ordinary bosons. 

For 2 and 2 particle ray exchanges:   For $p \geq 2$,
\ber
[ [ {a}^{\dag} ,{b}^{\dag} ], \{ {c}^{\dag} ,{d}^{\dag} \} ] =0, 
[ \{ {a}^{\dag} ,{b}^{\dag}\}, \{ {c}^{\dag} ,{d}^{\dag} \} ] =0, 
\eer
However, for $p= 2$, 
\ber
[ [ {a}^{\dag} ,{b}^{\dag} ],[  {c}^{\dag} ,{d}^{\dag} ] ] =0
\eer
but not zero for $p \geq 3$.   Hence, for $p\geq 3$ there is a ``non-commutative exchange" for both $ Y^{0}_2$ antisymmetric rays.   

Instead of (74, 75), when outside anticommutators, these all are not zero for $p \geq 2$.

For 1 and 3 particle ray exchanges:   For $p \geq 2$,
\ber
[ {a}^{\dag} , \{ {b}^{\dag} , {c}^{\dag} ,{d}^{\dag} \} ]  \neq 0, 
\{ {a}^{\dag} , \{ {b}^{\dag} , {c}^{\dag} ,{d}^{\dag} \} \} \neq 0,
\eer
For $p=1$, the first relation is zero.

For 2 and 3 particle ray exchanges, for $p$ arbitrary
\ber
[  \{ {a}^{\dag} ,{b}^{\dag}\}, \{ {c}^{\dag} , {d}^{\dag} ,{e}^{\dag} \} ] =0,  \nonumber
\{  \{ {a}^{\dag} ,{b}^{\dag}\}, \{ {c}^{\dag} , {d}^{\dag} ,{e}^{\dag} \} \} \neq 0
\eer
and for $p \geq 2$
\ber
[  [ {a}^{\dag} ,{b}^{\dag}], \{ {c}^{\dag} , {d}^{\dag} ,{e}^{\dag} \} ] \neq 0 
\eer
However, for $p= 2$, 
\ber
\{ [ {a}^{\dag} ,{b}^{\dag}], \{ {c}^{\dag} , {d}^{\dag} ,{e}^{\dag} \} \} =0,  
\eer
but not zero for $p \geq 3$.

For 3 and 3 particle ray exchanges,  for $p \geq 2$
\ber
[  \{ {a}^{\dag} ,{b}^{\dag},{c}^{\dag} \}, \{ {d}^{\dag}, {e}^{\dag} ,{f}^{\dag} \} ] \neq 0,  
\eer
but zero for $p=1$. When outside anticommutators, (79) is not zero for $p$ arbitrary.

Pairing of two parabose particles in two ``$p$-pairs" maybe important, in particular with respect to dark energy as it is in Bose-condensed systems:

Recall, for $ Y^{ l } _n = Y^{ 2 } _2 $  the 2-particle paraboson symmetric ray $ \{n_s, n_a\} = \{1, 0 \}$ and for $ Y^{ 0} _2 $ the antisymmetric ray $ \{n_s, n_a\} = \{0, 1 \}$.  Besides the above 2 paraboson ray exchange relations,  for the 2 paraboson symmetric ray for $p$ arbitrary 
\ber
[ \{ {a}^{\dag} ,{b}^{\dag} \}, c^{\dag}  d^{\dag}  \cdots {f_n}^{\dag} ] =0
\eer
for arbitrary $n=1, \cdots$ values.

For the 2 paraboson antisymmetric ray which does not occur for ordinary bosons 
and $m$-odd $\geq3$ ($n$-even $\geq 4$) particle ray exchanges, for $p =2$ 
\ber
\{  [ {a}^{\dag} ,{b}^{\dag}], \{ {c}^{\dag} , {d}^{\dag} , \cdots {f_m}^{\dag} \} \}=0,  \nonumber
\eer
\ber
[  [ {a}^{\dag} ,{b}^{\dag}], \{ {c}^{\dag} , {d}^{\dag} , \cdots {f_n}^{\dag} \} ] =0 
\eer
but not zero for $p \geq 3$. 
For opposite outside brackets, (81) are not zero for $p\geq 2$.

\subsection{\label{sec:level2} Parafermion Ray Exchanges }

For 1 and 2 particle ray exchanges:   For $p \geq 2$ 
\ber
[{\alpha}^{\dag}, [ {\beta}^{\dag} ,{\gamma}^{\dag}] ] =0, \{ {\alpha}^{\dag},[ {\beta}^{\dag} ,{\gamma}^{\dag}] \}\neq 0,  \nonumber
\eer
\ber
[ {\alpha}^{\dag}, \{ {\beta}^{\dag} ,{\gamma}^{\dag} \} ]\neq 0
\eer
For $p = 2$, 
\ber
\{ {\alpha}^{\dag}, \{ {\beta}^{\dag} ,{\gamma}^{\dag} \} \} = 0
\eer
but not zero for $p \geq 3$.   So, for $p\geq 3$ there is a ``non-commutative exchange" for the rays $\mathcal Y^{1}_1$ and $\mathcal Y^{0}_2$.   $ \mathcal Y^{0}_2$ is the 2 parafermion symmetric ray which does not occur for ordinary fermions. 

For 2 and 2 particle ray exchanges:   For $p \geq 2$,
\ber
[ \{ {\alpha}^{\dag} ,{\beta}^{\dag} \},[  {\gamma}^{\dag} ,{\delta}^{\dag} ] ] =0, 
[ [ {\alpha}^{\dag} ,{\beta}^{\dag}], [{\gamma}^{\dag} ,{\delta}^{\dag} ] ] =0, 
\eer
However, for $p= 2$, 
\ber
[ \{ {\alpha}^{\dag} ,{\beta}^{\dag} \},\{  {\gamma}^{\dag} ,{\delta}^{\dag} \} ] =0
\eer
but not zero for $p \geq 3$. Hence, for $p\geq 3$ there is  ``non-commutative exchange" for both $ \mathcal Y^{0}_2$ symmetric rays.  
Instead of (84, 85), when outside anticommutators, these all are not zero for $p \geq 2$.

For 1 and 3 particle ray exchanges:   For $p \geq 2$,
\ber
\{ {\alpha}^{\dag} , [ {\beta}^{\dag} , {\gamma}^{\dag} ,{\delta}^{\dag} ] \}  \neq 0, 
[ {\alpha}^{\dag} , [ {\beta}^{\dag} , {\gamma}^{\dag} ,{\delta}^{\dag} ] ] \neq 0,
\eer
For $p=1$, the first relation is zero.

For 2 and 3 particle ray exchanges, for $p$ arbitrary
\ber
[  [ {\alpha}^{\dag} ,{\beta}^{\dag}], [ {\gamma}^{\dag} , {\delta}^{\dag} ,{\epsilon}^{\dag} ] ] =0,  \nonumber
\{  [ {\alpha}^{\dag} ,{\beta}^{\dag} ], [ {\gamma}^{\dag} , {\delta}^{\dag} ,{\epsilon}^{\dag} ] \} \neq 0
\eer
and for $p \geq 2$
\ber
[  \{ {\alpha}^{\dag} ,{\beta}^{\dag} \}, [ {\gamma}^{\dag} , {\delta}^{\dag} ,{\epsilon}^{\dag} ] ] \neq0 
\eer
However, for $p= 2$, 
\ber
\{ \{ {\alpha}^{\dag} ,{\beta}^{\dag} \}, [ {\gamma}^{\dag} , {\delta}^{\dag} ,{\epsilon}^{\dag} ] \} =0,  
\eer
but not zero for $p \geq 3$.

For 3 and 3 particle ray exchanges,  for $p \geq 2$
\ber
  \{ [{\alpha}^{\dag} ,{\beta}^{\dag},{\gamma}^{\dag}], [ {\delta}^{\dag}, {\epsilon}^{\dag} ,{\zeta}^{\dag} ] \} \neq 0,  
\eer
but zero for $p=1$. When outside commutators, (89) is not zero for $p$ arbitrary.  

Pairing of two parafermion particles in two ``$p$-pairs" maybe important for dark energy, or otherwise, as it is for two fermions in Cooper-paired systems:
Besides the above 2 parafermion ray exchange relations,  for $ {\mathcal Y}^{ l } _n = \mathcal Y^{ 2 } _2 $ the 2 parafermion antisymmetric ray, $ \{\nu_a, \nu_s\} = \{1, 0 \}$ ,
 for $p$ arbitrary 
\ber
[  [ {\alpha}^{\dag} ,{\beta}^{\dag} ],  {\gamma}^{\dag}  {\delta}^{\dag}  \cdots {\eta_n}^{\dag}  ] =0
\eer
for arbitrary $n=1, \cdots$ values.  

For $ {\mathcal Y}^{ 0} _2$  the 2 parafermion symmetric ray $ \{\nu_a, \nu_s\} = \{0, 1 \}$ which does not occur for ordinary fermions,
and $m$-odd $\geq3$ ($n$-even $\geq 4$) particle ray exchanges, for $p =2$ 
\ber
\{ \{ {\alpha}^{\dag} ,{\beta}^{\dag}\} , [ {\gamma}^{\dag} , {\delta}^{\dag} , \cdots {\eta_m}^{\dag} ] \}=0,  \nonumber
\eer
\ber
[  \{ {\alpha}^{\dag} ,{\beta}^{\dag} \} , [ {\gamma}^{\dag} , {\delta}^{\dag} , \cdots {\eta_n}^{\dag} ] ] =0
\eer
but these do not vanish for $p \geq 3$.    For opposite outside brackets, (91) are not zero for $p\geq 2$.

\subsection{\label{sec:level2}Para-Family Ray Exchanges }

When paraparticles for $p \geq 2$ do not occur in a para-Family, then the parabosons and parafermions can be simply exchanged as one does for ordinary bosons and fermions, but this is not the case for paraparticles occurring in a para-Family.  In the case of a para-Family, for one pB and one pF there also exist two ``pB-pF mixed" generalized rays $ \{ {a}^{\dag},  {\beta}^{\dag} \}$  and $ [{a}^{\dag}, {\beta}^{\dag}] $  to consider the exchange properties of.  

\subsubsection{\label{sec:level2} For $p=2$}  

In the case of a para-Family, for 1 and 2 particle ray exchanges:  
\ber
\{ {a}^{\dag}, \{ {\beta}^{\dag} ,{\gamma}^{\dag} \}  \} =0,   [ {a}^{\dag},  [ {\beta}^{\dag} ,{\gamma}^{\dag} ] ] = 0,\nonumber 
\eer
\ber
\{  {\alpha}^{\dag}, [ {b}^{\dag} ,{c}^{\dag} ] \} = 0,  [ {\alpha}^{\dag},\{  {b}^{\dag} ,{c}^{\dag} \} ] = 0
\eer
and for the two ``pB-pF mixed" two particle rays
\ber
[ {a}^{\dag},\{  {b}^{\dag} ,{\gamma}^{\dag} \} ] = 0, \{  {a}^{\dag}, [ {b}^{\dag} ,{\gamma}^{\dag} ] \} = 0, \nonumber
\eer
\ber
\{ {\alpha}^{\dag}, \{ {b}^{\dag} ,{\gamma}^{\dag} \}  \} =0, [ {\alpha}^{\dag},  [ {b}^{\dag} ,{\gamma}^{\dag} ] ] = 0
\eer
For opposite outside brackets, these are not zero.

For 2 and 2 particle ray exchanges:  
\ber
[ [ {a}^{\dag} ,{b}^{\dag} ], \{ {\gamma}^{\dag} ,{\delta}^{\dag} \} ] =0, 
[ [ {a}^{\dag} ,{b}^{\dag} ], [ {\gamma}^{\dag} ,{\delta}^{\dag} ] ] =0 \nonumber
\eer
\ber
[ \{{a}^{\dag} ,{b}^{\dag} \}, \{ {\gamma}^{\dag} ,{\delta}^{\dag} \} ] =0, 
[ \{ {a}^{\dag} ,{b}^{\dag}\}, [ {\gamma}^{\dag} ,{\delta}^{\dag} ]  ] =0, 
\eer

For all  ``pB-pF mixed"  particle rays
\ber
\{ [ {a}^{\dag} ,{\beta}^{\dag} ], [ {c}^{\dag} ,{\delta}^{\dag} ] \}=0,
\{ \{ {a}^{\dag} ,{\beta}^{\dag}\}, \{ {c}^{\dag} ,{\delta}^{\dag} \}  \}=0,  \nonumber
\eer
\ber
\{ \{ {a}^{\dag} ,{\beta}^{\dag}\}, [ {c}^{\dag} ,{\delta}^{\dag} ]  \}=0, 
\eer

For mixed with 2 parabosons
\ber
[ [ {a}^{\dag} ,{\beta}^{\dag} ], [ {c}^{\dag} ,{d}^{\dag} ] ] =0, 
[ \{ {a}^{\dag} ,{\beta}^{\dag} \}, \{ {c}^{\dag} ,{d}^{\dag} \} ] =0 \nonumber
\eer
\ber
[  \{ {a}^{\dag} ,{\beta}^{\dag} \}, [ {c}^{\dag} ,{d}^{\dag} ] ] =0, 
[ [ {a}^{\dag} ,{\beta}^{\dag} ],  \{ {c}^{\dag} ,{d}^{\dag} \} ]  =0
\eer

For mixed with 2 parafermions
\ber
[ [ {a}^{\dag} ,{\beta}^{\dag} ], \{ {\gamma}^{\dag} ,{\delta}^{\dag} \} ] =0, 
[ [ {a}^{\dag} ,{\beta}^{\dag} ], [ {\gamma}^{\dag} ,{\delta}^{\dag} ] ] =0 \nonumber
\eer
\ber
[ \{ {a}^{\dag} ,{\beta}^{\dag}\}, \{{\gamma}^{\dag} ,{\delta}^{\dag}\} ] =0, 
[ \{ {a}^{\dag} ,{\beta}^{\dag}\}, [ {\gamma}^{\dag} ,{\delta}^{\dag} ]  ] =0, 
\eer
For opposite outside brackets, these are not zero

For $p=2$, in the case of a para-Family the pairing of two quanta in two ``$p$-pairs" has vanishing trilinear relations 
\ber
[ {\rho}^{\dag},[  {\sigma}^{\dag} ,{t}^{\dag} ] ] = 0, \{ {r}^{\dag},\{  {\sigma}^{\dag} ,{\tau}^{\dag} \} \} = 0
\eer
which are not zero for $p \geq 3$.  So for $p=2$, the pB-pF commutator pair commutes with any number of pF operators
\ber
[ [ {\rho}^{\dag} ,{s}^{\dag} ], {\tau}^{\dag}  {\upsilon}^{\dag}  \cdots {{\omega}_n}^{\dag} ] =0.
\eer
The pair brackets inside these 3 relations vanish for ordinary bosons and fermions.

Relations for ``$p$-pairs" for $p$ arbitrary are in the next subsection.

\subsubsection{\label{sec:level2} For $p \geq 3$} 

In the case of a para-Family, for 1 and 2 particle ray exchanges:  There are the fundamental trilinear relations
\ber
   [ {a}^{\dag},  [ {\beta}^{\dag} ,{\gamma}^{\dag} ] ] = 0, [  {\alpha}^{\dag}, \{ {b}^{\dag} ,{c}^{\dag} \} ] = 0
\eer
However,
\ber
\{ {a}^{\dag}, \{ {\beta}^{\dag} ,{\gamma}^{\dag} \} \} \neq 0, \{  {\alpha}^{\dag}, [ {b}^{\dag} ,{c}^{\dag} ] \} \neq 0
\eer
For the two ``pB-pF mixed" two particle rays, the tri-linear relations are
\ber
[ {a}^{\dag},\{  {b}^{\dag} ,{\gamma}^{\dag} \} ] = 0, \{  {\alpha}^{\dag}, \{ {b}^{\dag} ,{\gamma}^{\dag} \} \} = 0
\eer
However,
\ber
\{ {a}^{\dag},[ {b}^{\dag} ,{\gamma}^{\dag}]  \} \neq 0, [  {\alpha}^{\dag}, [ {b}^{\dag} ,{\gamma}^{\dag} ] ] \neq 0
\eer

For 2 and 2 particle ray exchanges:  
\ber
[ [ {a}^{\dag} ,{b}^{\dag} ], [ {\gamma}^{\dag} ,{\delta}^{\dag} ] ] =0, 
[ \{ {a}^{\dag} ,{b}^{\dag} \},\{ {\gamma}^{\dag} ,{\delta}^{\dag} \} ] =0 \nonumber
\eer
\ber 
[ \{ {a}^{\dag} ,{b}^{\dag}\}, [ {\gamma}^{\dag} ,{\delta}^{\dag} ]  ] =0, 
\eer

For all  ``pB-pF mixed"  particle rays
\ber
\{ \{ {a}^{\dag} ,{\beta}^{\dag}\}, \{ {c}^{\dag} ,{\delta}^{\dag} \}  \}=0,
\{ \{ {a}^{\dag} ,{\beta}^{\dag}\}, [ {c}^{\dag} ,{\delta}^{\dag} ]  \}=0,  
\eer

For mixed with 2 parabosons
\ber
[ \{ {a}^{\dag} ,{\beta}^{\dag} \}, \{ {c}^{\dag} ,{d}^{\dag} \} ] =0, 
[  \{ {a}^{\dag} ,{\beta}^{\dag} \}, [ {c}^{\dag} ,{d}^{\dag} ] ] =0 \nonumber
\eer
\ber
[ [ {a}^{\dag} ,{\beta}^{\dag} ],  \{ {c}^{\dag} ,{d}^{\dag} \} ]  =0
\eer

For mixed with 2 parafermions
\ber
[ [ {a}^{\dag} ,{\beta}^{\dag} ], [ {\gamma}^{\dag} ,{\delta}^{\dag} ] ] =0 
[ \{ {a}^{\dag} ,{\beta}^{\dag}\}, \{{\gamma}^{\dag} ,{\delta}^{\dag}\} ] =0, 
 \nonumber
\eer
\ber
[ \{ {a}^{\dag} ,{\beta}^{\dag}\}, [ {\gamma}^{\dag} ,{\delta}^{\dag} ]  ] =0, 
\eer

However,
\ber
[ [ {a}^{\dag} ,{b}^{\dag} ], \{ {\gamma}^{\dag} ,{\delta}^{\dag} \} ] \neq 0, 
[ [ {a}^{\dag} ,{b}^{\dag} ], [ {c}^{\dag} ,{\delta}^{\dag} ] ]  \neq 0 \nonumber
\eer
\ber
[\{ {\alpha}^{\dag} ,{\beta}^{\dag}  \}, [ {c}^{\dag} ,{\delta}^{\dag} ]  ]  \neq 0, 
\{  [{a}^{\dag} ,{\beta}^{\dag} ], [ {c}^{\dag} ,{\delta}^{\dag} ] \}\neq 0
\eer

For $p$ arbitrary, in the case of a para-Family in the pairing of two quanta in two ``$p$-pairs" there are vanishing relations for the non-zero pairs which occur for ordinary $p=1$ particles
\ber
[ \{ {\rho}^{\dag} ,{s}^{\dag} \}, t^{\dag}  u^{\dag}  \cdots {w_n}^{\dag} ] =0, \nonumber
\eer
\ber
[ \{ {r}^{\dag} ,{s}^{\dag} \},{\tau}^{\dag}  {\upsilon}^{\dag}  \cdots {{\omega}_n}^{\dag} ] =0,  \nonumber
\eer
\ber
[ [ {\rho}^{\dag} ,{\sigma}^{\dag} ],{t}^{\dag}  {u}^{\dag}  \cdots {{w}_n}^{\dag} ] =0
\eer
for arbitrary $n=1, \cdots$ values.

For the 2 paraparticle rays which do not occur for $p=1$ particles:  For $p\geq 3$ the trilinear relation 
\ber
\{ \{ {\rho}^{\dag}, {s}^{\dag} \} ,{\tau}^{\dag}  \} = 0, \nonumber
\eer
is equivalent to
\ber
[ [ {s}^{\dag}, {\tau}^{\dag} ] ,{\rho}^{\dag} ] = \{ \{ {\tau}^{\dag}, {\rho}^{\dag} \} ,{s}^{\dag}  \}
\eer
which is zero for $p=2$, but is not zero for $p \geq 3$. 

\section{Summary Remarks}

Parastatistics requires the bracket generalized ray representatives constructed in this paper for parabosons, parafermions, and for possible para-Families be used for specifying the asymptotic states in perturbative S-matrix calculations to determine whether or not paraparticles are associated with dark matter ( $\sim 100$ GeV to $\sim 40$ TeV mass scale).   

In the construction of the portal Lagrangian densities in [5], two pF fields occur in a commutator ordering.  Two pB fields, or a pB and a pF field in a pFam, occur in an anti-commutator ordering.  This occurs because it is assumed that ``second unit observables" do not occur.  See Appendix B of [5].    

Consequently, in calculations of cross-sections for the annihilation of two dark matter paraparticles there will be an extra
  ``effective-flux-factor" of $\frac{1}{2}$.  This factor occurs due to the decoupling in the scattering amplitude of the other asymptotic initial state.   In the alternate case of commuting pB and pF fields this suppression factor is absent.  Therefore, the presence or absence of this factor is a simple empirical test for the existence of parastatistics and a test for the occurrence of a para-Family, versus commuting pB's and pF's. 

The analogous decoupling factor/test occurs in the production of two paraparticles as one generalized final ray is coupled and the other is not.   The symmetric (asymmetric) ray is coupled (not coupled) in the case of two pB's, and a pB and a pF in a pFam.  The asymmetric ray is the coupled one in the case of two pF's.

Determination of the presence/absence of this extra factor will also provide fundamental quantum-field-theoretic locality information:  

 For $p=2$, if two pB (pF) currents commute when sufficiently space-like separated, there is ``weak locality" with the occurrence of second units.  However, if instead the field and the current commute, there is ``strong locality" and the absence of second units.  See [2, 4] and Appendix B in [5].  

If paraparticles are less massive  ($\sim 0.01$ eV, neutrino mass scale) and associated with dark energy, via a condensate or otherwise, the occurrence of both types of ``$p$-pairs" may be dynamically important even if one of the pairs is not coupled directly in the interaction Lagrangian densities.   Due to the $ \overline{P}_{i,j} $ particle exchange symmetry, in a quantum domain of more than one mode the paraparticles must be treated in generalized rays. 

In all cases, the generalized rays occur in more inclusive ray multiplets.   In the case of para-Families, the four Kinds of pFam's each have their ray multiplets naturally occurring in an analogous Meta-Multiplet.   

For a large $n$ number of pB's, the ``most probable ray"  $Y^{l_{mpr}}_n$ has $l_{mpr} = \sqrt{n}$, is ray $r_{mpr} \sim \frac{n}{2} \sim r_{max}$, and is narrow in $r$.   For a large number $\nu$ of pF's, the ``most probable ray"   $ {\mathcal Y}^{{\lambda}_{mpr}}_\nu $, has $\lambda_{mpr} = \sqrt{\nu}$, is ray $\rho_{mpr} \sim \frac{\nu}{2} \sim \rho_{max}$, and is narrow in $\rho$.

The explicit arbitrary order $p$ ray representatives for a few paraparticles, see Section V, may be useful to establish the order $p$ if dark matter and/or dark energy is discovered to exhibit parastatistic properties.

 The ``exchange relations for the brackets" of Section VI may be useful in the analysis of a decay or scattering process with two final rays in different regions in coordinate space.    Depending on whether $p = 2$ or $p \geq 3$, the two generalized ray operators often commute or don't commute.  
 
 Neutral dark matter and/or dark energy paraparticles might occur with exact, or approximate, supersymmetry as in the ``statistics portal"  Wess-Zumino model with spin-0 particle $\breve A$ and antiparticle  $\breve{B}= \overline{\breve{A}}$  plus a spin-1/2 para-Majorana neutrino  $ \breve{\xi}=\breve{\nu}$, see [5].     

\begin{acknowledgments}
We thank Eric Aspling for useful discussions.
\end{acknowledgments}

\appendix

\section{Fundamental Trilinear Commutation Relations }

In the parastatistics literature, see [2,4], a useful distinction is made between the general ``trilinear commutation relations" (TCR) for arbitrary $p$ values and the stronger ``self-contained commutation relations" (SCR) which hold for specific $p$ values, $p=1,2, 3$.   For $p=1$ the SCR are the usual boson and fermion bilinear relations.  For $p=2$, the trilinear SCR for $p=2$ are listed in Appendix A of [5]  for parabosons, parafermions, and for the case of a para-Family.   The ``flip" reordering relations used in Section II only hold for two SCR cases $p=1,2$.  For $p=3$, the lengthy SCR for pB's and pF's are listed in Appendix B of [4].

For ordinary bosons and fermions, the relative signature of the two fields [2, 4] is normally chosen as $+1$, so their field operators commute.  In [5] for $p=2$, and in the present paper for $p$ arbitrary, the relative signature for pB's and pF's is chosen as $+1$ for both the direct product case $ \overline{Y}^{  } _n  \times  \overline{\mathcal Y}^{   } _\nu $ and for the case of a para-Family  
$ \overline{Y}^{  } _n \overline{\mathcal Y}^{  } _\nu$.

To write compactly the TCR and SCR for parastatistics, we define ``ordered Kronecker delta's" which have ordered, operator-subscripts.  Like the usual Kronecker delta, they commute.

They can be used for the ordinary boson and fermion commutation relations, see (A3) below. 

When both subscripts are for the same, so $a=b$ or $\alpha = \beta$, these  ``ordered Kronecker delta's" do not vanish, instead
\ber
\cancel{\delta}_{a {b}^{\dag}} =  - \cancel{\delta}_{{b}^{\dag} a }= 1, \hat{\delta}_{a {b}^{\dag}} =    \hat{\delta}_{ {b}^{\dag} a}=1 ~ (a=b)  \nonumber
\eer
\ber
\cancel{\delta}_{\alpha {\beta}^{\dag}}= - \cancel{\delta}_{ {\beta}^{\dag}\alpha}=1, \hat{\delta}_{\alpha  {\beta}^{\dag}}= \hat{\delta}_{  {\beta}^{\dag}\alpha}=1  ~ (\alpha = \beta)
\eer
Like the usual Kronecker delta, these vanish otherwise.  These also vanish when both subscripts are ``dagger" or both ``no-dagger."

The ``slash" accent versus the ``hat" accent on these ordered Kronecker delta's is to denote the minus signs in (A1).  The ``slash" ones are antisymmetric in indices; the ``hat" ones are symmetric.  Note that the positive sign for the ``slash" ones has been chosen for when the creation and annihilation operators are in the order of the non-vanishing paraparticle vacuum condition 
\ber
a_k a_l^{\dag}  | 0 > =   {\alpha}_k {\alpha}_l^{\dag}                      | 0 > = p \, \delta_{kl}  | 0 > 
\eer
For the para-Majorana operators the usual mode Kronecker delta is $ \delta_{lm}=\delta_{\lambda_l \lambda_m}  \delta^{(3)} (\vec {p}_l - \vec{p}_m)$ with ${\lambda_l} $ and ${\lambda_m}$ helicity indices, and analogously for the para-Boson operators.   

Two lines of equations are written in (A1) to stress that this sign convention is the same for pB operators (Roman symbols) and for pF operators (Greek symbols).

Below, a ``hat" on a creation or annihilation operator symbol, $\hat a$ or $ \hat{\alpha}$,  denotes that the equation holds for both the case when the operator has (hasn't) a ``dagger,"  throughout the equation.  Below, the ``dagger,"  ``no-dagger,"  or ``hat" type accent on the subscripts in (A1) will be omitted.   The specific accent is to be understood to go with the Roman or Greek letter, through out the equation, including the operator-subscripts on these ordered Kronecker delta's.

In this compact notation, the SCR for ordinary $p=1$ boson and fermion operators are 
\ber
[ \hat{r},  \hat{s} ] = \cancel{\delta}_{r s} , \{  \hat{{\rho}},  \hat{{\sigma}} \} =  \hat{\delta}_{\rho \sigma}, [ \hat{r},  \hat{\sigma} ] =0.
\eer

\subsubsection{\label{sec:level2} Trilinear Commutation Relations in  Compact Form}

For arbitrary order $p$, in compact form the TCR for all parabosons are
\ber
[ \hat{r}, \{ \hat{s} ,\hat{t} \} ] = 2 \cancel{\delta}_{r s} \hat{t} + 2 \cancel{\delta}_{r t} \hat{s}
\eer
and for all parafermions
\ber
[ \hat{{\rho}}, [ \hat{{\sigma}} ,\hat{{\tau}}] ] = 2\hat{\delta}_{\rho \sigma} \hat{\tau} - 2 \hat{\delta}_{\rho \tau} \hat{\sigma}
\eer

For $p=2$ order, there are the extra relations: for all parabosons
\ber
\{ \hat{r}, [ \hat{s} ,\hat{t} ] \} = 2 \cancel{\delta}_{r s} \hat{t} - 2 \cancel{\delta}_{r t} \hat{s} + 4 \cancel{\delta}_{s t} \hat{r}
\eer
or the simpler ``flip" SCR 
\ber
 \hat{r}  \hat{s}  \hat{t} -   \hat{t}  \hat{s}  \hat{r}  = 2 \cancel{\delta}_{r s} \hat{t}  + 2 \cancel{\delta}_{s t} \hat{r}
\eer
Similarly, for all parafermions there are the extra relations
\ber
\{ \hat{{\rho}},\{ \hat{{\sigma}} ,\hat{{\tau}} \} \} = 2\hat{\delta}_{\rho \sigma} \hat{\tau} + 2 \hat{\delta}_{\rho \tau} \hat{\sigma} +4  \hat{\delta}_{\sigma \tau} \hat{\rho}
\eer
or the ``flip" SCR 
\ber
 \hat{\rho} \hat{\sigma} \hat{\tau}  +  \hat{\tau} \hat{\sigma} \hat{\rho} = 2 \hat{\delta}_{\rho \sigma} \hat{\tau} + 2 \hat{\delta}_{\sigma \tau} \hat{\rho}
\eer

For a para-Family, in compact form the TCR are for two parabosons and one parafermion
\ber
[ \hat{r}, \{ \hat{\sigma} ,\hat{t} \} ]  = 2 \cancel{ \delta}_{r  t}  \hat{\sigma}, [ \hat{\rho}, \{ \hat{s} ,\hat{t} \} ] = 0
\eer
For two parafermions and one paraboson,
\ber
\{ \hat{\rho}, \{ \hat{\sigma} ,\hat{t}  \} \} = 2 \hat{ \delta}_{\rho  \sigma}  \hat{t},
 [ \hat{r}, [ \hat{\sigma} ,\hat{\tau} ] ]= 0
\eer

For a para-Family, for $p=2$ order, there are the extra relations: for two parabosons and one parafermion
\ber
\{ \hat{\rho}, [  \hat{s} ,\hat{t} ] \} =4 \cancel{ \delta}_{s  t}  \hat{\rho}, \{ \hat{r}, [  \hat{\sigma} ,\hat{t} ] \} =  - 2 \cancel{ \delta}_{r  t}  \hat{\sigma},
\eer
or the ``flip" SCR 
\ber
 \hat{\rho}  \hat{s} \hat{t} -  \hat{t}  \hat{s}   \hat{\rho} = 2 \cancel{ \delta}_{s  t}  \hat{\rho} , \hat{r} \hat{\sigma}  \hat{t} -   \hat{t} \hat{\sigma}  \hat{r}= 0
\eer
For two parafermions and one paraboson, the extra relations are
\ber
[ \hat{\rho}, [ \hat{\sigma} ,\hat{t} ] ] = 2 \hat{ \delta}_{\rho  \sigma}  \hat{t}, \{ \hat{r}, \{ \hat{\sigma} ,\hat{\tau} \} \} = 4 \hat{ \delta}_{  \sigma \tau}  \hat{r}
\eer
or the ``flip" SCR 
\ber
 \hat{\rho} \hat{\sigma}  \hat{t} +   \hat{t}  \hat{\sigma} \hat{\rho} = 2 \hat{ \delta}_{\rho  \sigma}  \hat{t},
 \hat{\rho}  \hat{s}  \hat{\tau} +   \hat{\tau}  \hat{s} \hat{\rho} = 0
\eer

The usage of these ``ordered Kronecker delta's" in writing compactly these TCR and SCR relations shows them to be basic quantities in the quantized field treatment of paraparticles.   

\subsubsection{\label{sec:level2} Trilinear Commutation Relations 
\newline for $p$ Arbitrary in Conventional Form}

The SCR for $p=2$, without using ordered Kronecker delta's, are listed in Appendix A of [5].
The corresponding TCR for $p$ arbitrary are the following:

For all pB's:  
\ber
[ r, \{ s ,t \} ] = 0, [  \{ r, s \} , t^{\dag}] = 2 \delta_{st} r + 2 \delta_{rt} s, \nonumber
\eer
\ber
 [ r, \{ s^{\dag} ,t \} ] = 2 \delta_{r s} t
 \eer

For all pF's:
\ber
[ {\rho}, [ {\sigma} ,{\tau}] ] = 0,[  [ {\rho} ,{\sigma}] , \tau^{\dag}] = 2 \delta_{\sigma \tau} \rho - 2 \delta_{\rho \tau} \sigma, \nonumber
\eer
\ber
[ \rho, [ \sigma^{\dag} ,\tau ]] = 2 \delta_{\rho \sigma} \tau
\eer

For a para-Family, for two pB's and one pF:
\ber
[ r, \{ \hat{\sigma} ,t \} ] = 0, [   \{ r ,{\sigma} \}  , t^{\dag}] = 2 \delta_{r t}  \sigma, \nonumber
\eer
\ber
 [ \hat{\rho}, \{ \hat{s} ,\hat{t}  \} ] = 0
 \eer
  
For a para-Family, for two pF's and one pB:
\ber
\{{\rho}, \{{\sigma} ,t \} \} = 0,  \{ \rho, \{ \sigma^{\dag}, \hat{t} \}  \} = 2 \delta_{\rho \sigma} \hat{t}, \nonumber
\eer
\ber
[ \hat{r},  [ \hat{ \sigma} ,\hat{ \tau} ] ] = 0
\eer

\section{Fundamental Green Indices}

In parastatistics, both the general TCR for arbitrary order $p$, the SCR for a specific $p$ order, and the states in the generalized ray representatives can be written without the use of Green indices.   The $p=2$ results in this paper do not require the use of Green indices.  

However, any matrix element of creation and annihilation operators, between states in generalized rays, has an explicit value determined by the use of Green indices.   The same is true for vacuum expectation values of any collection of parafields.   The ``exchange relations for the brackets'' treated in Section V are straight-forwardly determined by use of the Green indices.   By Green indices, it is easy to see that the ``flip" reordering relations only hold for $p=1,2$.  While consistency requirements, such as for higher order nested commutators and/or for orthonormality of the states in generalized rays for arbitrary $p$, might yield the same results, it is simpler to just view Green indices as fundamental in the formulation parastatistics.  They are used in extending the proofs of fundamental theorems from $p=1$ quantum field theory to $p \geq 2$ [2, 9, 8, 4].

Nevertheless for operator simplifications, caution is needed:  While it is always possible to expand a operator-valued function of creation and annihilation operators in Green indices and then proceed to simplify it, i.e. writing it with fewer Green indices, it frequently occurs that if the result is non-zero, it's expression without using Green indices is not apparent, other than by the expression which was expanded.

The Green component fields are $  {\breve { \phi} ^{(a)}_i (x)}$ defined by the expansion
\ber
\breve { \phi} _i (x) = { \sum_{a=1}^p } \;  { \breve { \phi}^{(a)}_i (x)}.
\eer
where the superscript, lower-case Roman-letter $a$ is the Green index.   The analogous expansion is made for a creation or annihilation operator. For pB's (pF's) two Green component fields $  {\breve { \phi} ^{(a)}_i (x)}, {\breve { \phi} ^{(b)}_j (x)}$ with the same Green index, $a=b$, 
obey the usual Bose (Fermi) commutation relations, but  anticommute (commute) when their two Green indices are different, $a \neq b$.  In an ``order $p$ family" of parafields, a parabose Green component and a parafermi Green component have the same commutation pattern as two paraboson Green components.   There are, of course, no contractions between the commuting pB and pF Green component operators.

Because paraparticles may occur in a pFam, it is best to use the same symbol for superscript Green indices for pB's and pF's.  We use lower-case Roman-letters $a, b \cdots$ for Green indices to avoid confusion with our use of capital Roman-letters $A, B \cdots $  indices (dotted and undotted) for two-component Weyl spinor fields $\breve{\xi}_A (x)$ [5]. 
 
The vacuum condition for Green components 
\ber
a^{(a)}_k | 0 > = \alpha^{(a)}_l |0> =0 , \; <0|0> =1    \nonumber \\
a^{(a)}_k {a^{(a)}_l}^{\dag}  | 0 > =   {\alpha}^{(a)}_k {{\alpha}^{(a)}}_l^{\dag}                      | 0 > =  \, \delta_{kl}  | 0 >  
\eer
gives for mode operators with indices $k, l $ 
\ber
a_k | 0 > = \alpha_l |0> =0      \nonumber \\
a_k a_l^{\dag}  | 0 > =   {\alpha}_k {\alpha}_l^{\dag}                      | 0 > = p \, \delta_{kl}  | 0 >  
\eer
 and $ a_k  {\alpha}_l^{\dag} | 0 > =   {\alpha}_k       a_l^{\dag}                | 0 > =0$.   In (B2), the Green index is not summed.  For pF operators there is the single-mode constraint $  ( {\alpha}^{\dag} )^{p + 1} = 0 $, but there is none for pB operators.  The parabose and parafermi number mode operators are 
  $ N_k= \frac{1}{2} \{ a_k^{\dag}, a_k \}  - p/2   $  and  $
\mathcal{ N}_k= \frac{1}{2} [ \alpha_k^{\dag}, {\alpha}_k ] +  p/2$.  At most $p$ identical pF's (pB's) can occupy a totally antisymmetric (symmetric) state.  

These important $p$ dependences can be simply viewed as coming from the Green indices. 

These Hermitian number operators satisfy $[ a_k, N_l ]= a_k \delta_{k l}, [ \alpha_k, \mathcal{ N}_l ]= \alpha_k \delta_{k l}$ with $[ N_k, N_l ]=  [ N_k, \mathcal{ N}_l ]=  [ \mathcal{N}_k, \mathcal{ N}_l ] = [ a_k, \mathcal{N}_l ]=[ \alpha_k, N_l ]=0$.  These hold for both the generalized ray multiplet's case $ \overline{Y}^{  } _n  \times  \overline{\mathcal Y}^{   } _\nu $ and for the case of a pFam 
$ \overline{Y}^{  } _n \overline{\mathcal Y}^{  } _\nu$.

\end{document}